\documentclass[%
 reprint,
groupedaddress,
frontmatterverbose, 
showpacs,preprintnumbers,
 amsmath,amssymb,
 aip,
 jcp,
floatfix
]{revtex4-1}

\usepackage{graphicx,url,graphics,epstopdf}
\usepackage{dcolumn}
\usepackage{bm}
\usepackage{longtable}                                                                           \usepackage[dvipsnames]{xcolor}
\begin{document}
\title{ Towards a spectroscopically accurate set of potentials for heavy hydride laser cooling candidates: effective core potential calculations of BaH}

\author{Keith Moore}

\author{Brendan M.\ McLaughlin}

\author{Ian C.\ Lane} \email{i.lane@qub.ac.uk}
\affiliation{School of Chemistry and Chemical Engineering, Queen's University Belfast, Stranmillis Road, Belfast BT9 5AG, UK}

\date{\today}

\begin{abstract}
BaH (and its isotopomers) is an attractive molecular candidate for laser cooling to ultracold temperatures and a potential precursor 
for the production of ultracold gases of hydrogen and deuterium. The theoretical challenge is to simulate the 
laser cooling cycle as reliably as possible and this paper addresses the generation of a highly accurate 
{\it ab initio} $^{2}\Sigma^+$ potential for such studies. The performance of various basis sets within the 
multi-reference configuration-interaction (MRCI) approximation with the Davidson correction (MRCI+Q)
is tested and taken to the Complete Basis Set (CBS) limit. 
It is shown that the calculated molecular constants using a 46 electron Effective Core-Potential (ECP) and even-tempered augmented polarized core-valence basis sets (aug-pCV$n$Z-PP, $n$ = 4 and 5) but only including three 
active electrons in the MRCI calculation are in excellent agreement with the available experimental values. 
The predicted dissociation energy D$_e$ for the X$^2\Sigma^+$  state (extrapolated to the CBS limit)
 is 16895.12 cm$^{-1}$ (2.094 eV), which agrees within 0.1$\%$ of a revised experimental value of $<$16910.6 cm$^{-1}$, while the calculated r$_e$ is within 0.03 pm of the experimental result.
\end{abstract}


\maketitle

\section{Introduction}
\noindent
Recently, it has been proposed \cite{Lane2015} that ultracold hydrogen atoms may be formed by the photo-dissociation of laser cooled hydrides, provided the energy 
of the fragmentation products is much smaller than the average thermal energy of the parents. At the limit of zero kinetic 
energy release, the velocities of the hydrogen atoms will match that of the parents, forming a Maxwellian velocity 
distribution corresponding to a much lower hydrogen atom temperature than the original molecular gas (T$_H$ = $\frac{T_{MH}}{m_{MH}}$). 
Quantum chemical calculations of the transition dipoles and Franck-Condon (FC) factors, using the post Hartree-Fock
 methods Complete Active Space Self Consistent Field (CASSCF) \cite{Knowles1985}
 and MRCI \cite{Knowles1988}, confirmed that BaH was a very good
 candidate for demonstrating this kinematic effect as the barium atom has a considerable mass and the molecular radical is amenable to laser cooling\cite{Tarallo2016}. In the earlier study a very small (triple zeta quality) 
two electron basis set for barium was used with the inner 52 electrons replaced by an ECP. To improve agreement with
 the experimental energy levels of the atom, an $\ell$-independent Core-Polarization Potential (CPP) was added to the calculation. 
Although the resulting basis set simulated the reported experimental FC factors very well, there were clearly issues 
with the calculation of the ground state equilibrium bond length r$_e$ (experimental value\cite{Bernath2013} 2.23188651(19) $\text{\AA}$) and the dissociation limit\cite{Preuss1987,Kaupp1991,Allouche1992,Skripnikov2013}. This paper attempts to produce a global, high-quality {\it ab initio} potential for the X$^2\Sigma^+$ ground state of BaH. The heavier familial hydrides YbH and RaH are also attractive precursors of ultracold hydrogen but, unlike BaH, there is currently a lack of reliable theoretical or experimental data to confirm their suitability. Hydrogen forms the basis of many spectroscopic tests of fundamental symmetries\cite{Kostelecky2015}, for the evolution of the Universe\cite{Ubachs2016} and in metrology\cite{Rigden1983,Kragh1985} (for example, the measurement of the Rydberg constant\cite{Cagnac1994} and the proton radius\cite{Karshenboim2015}).

BaH is naturally a heavier cousin of BeH, the simplest stable neutral open shell diatomic. BeH possesses just five electrons and 
has recently been the focus of two important papers \cite{Koput2011,Dattani2015} regarding the simulation of its ground potential energy curve. 
The first\cite{Koput2011} was a detailed {\it ab initio} study using very large correlation consistent basis sets (aug-cc-pCV$n$Z)
 up to seven zeta ($n$ = 7)  in character (indeed so large that the MOLPRO\cite{Werner2010} {\it ab initio} package could not use 
the largest $\ell$ = 7 functions in the basis set 
and the effect of excluding them was determined by a separate calculation). When determining the static electron correlation
 beyond Hartree-Fock level, the CASSCF wavefunction included the excited $3s$ and $3p$ 
orbitals on the Be atom within the active 
space despite this being a single electronic state calculation. 
To avoid the size-extensivity issues inherent in MRCI (for dynamic electron correlation) 
the MR-ACPF \cite{Gdanitz1988}(multi-reference averaged coupled-pair functional) method was used instead. The size-extensivity and 
basis set superposition errors (BSSE) were determined to both be around 1-2 cm$^{-1}$ at the $n$ = 7 level of calculation. 
The calculations were then extrapolated to the CBS limit. Despite the meticulous care taken in producing this 
{\it ab initio} potential, the resulting equilibrium bond length deviated by {0.13~pm} from the spectroscopic value. 
While in excellent agreement for a typical {\it ab initio} calculation, for a benchmark molecule it is a little disappointing 
that such a powerful calculation could not achieve a better agreement. Tremendous pains were then made\cite{Koput2011} to include 
a host of minor corrections to the potential including electron correlation beyond the MR-ACPF method (to a level equivalent to full CI), scalar relativistic 
effects and diagonal Born-Oppenheimer corrections (DBOC). However, the energy effect of each of these additional contributions 
(all calculated using much smaller basis sets than used in the MRCI calculation) was small and mutually cancelling, while the improvement in bond length was just 0.04 pm. Furthermore, the bond dissociation energy was 90 cm$^{-1}$ 
higher than the best experimental data \cite{Leroy2006} at that point  (17590 $\pm$ 200 cm$^{-1}$), though smaller than the raw 
{\it ab initio} result (17699 cm$^{-1}$). The second paper \cite{Dattani2015} concerned taking that experimental work by Le Roy {\it et al} \cite{Leroy2006} and 
fitting a new analytical function to the spectroscopic data. Just 5\% of the rotationless adiabatic potential has 
yet to be covered experimentally and still the quantum number of the highest bound vibrational state is unclear. 
Again great attention to detail was conducted in this study, including mass corrected $C_6$, $C_8$ and $C_{10}$ 
constants for the different hydrogen isotopes. Despite this, the question of whether $v$ = 12 is the highest vibrational 
level is still unresolved but the experimental D$_e$ is now within 1 cm$^{-1}$ of the corrected {\it ab initio} value. If confirmed by future experiments, 
this would be an amazing achievement for {\it ab initio} quantum chemistry.

Given the difficulties in describing the relatively simple BeH, the task of completing accurate potentials for the heavier system BaH, 
that contains an additional 52 electrons, looks formidable. Furthermore, relativistic effects that were tiny in BeH are amplified in such 
a heavy hydride. However, 46 or even 54 of the electrons can be replaced with an ECP and many of the scalar relativistic effects 
can be included in that potential. The larger core was used by Lane \cite{Lane2015} and earlier by Preuss and co-workers \cite{Preuss1987} but in both instances 
the predicted ground state bond length was calculated to be much too short using a triple zeta basis set.
 While Preuss and co-workers \cite{Preuss1987,Kaupp1991} managed to achieve 1 pm accuracy 
using a smaller ($4s$,$4p$,$2d$) basis set, such behavior is symptomatic 
of simple error correction. Furthermore, both D$_e$ and $\omega_e$ were 
much lower \cite{Preuss1987} than experimental estimates at that time. It should not
 be forgotten that, despite the short bond length, the 2-electron, CPP + ECP calculation of FC 
factors is sufficiently accurate \cite{Lane2015} to be useful for screening good laser cooling transitions.
 For accurate calculation of the molecular constants, however, the evidence points 
to a minimum ten-electron basis set for barium.

Adopting the smaller 46-electron ECP allows the introduction of core-valence interactions with the $5s$ and $5p$ electrons as well as providing 
an opportunity to adjust the number of electrons used in calculating static and dynamic electron correlation. The change in ECP increased 
the predicted  $r_e$ value of Kaupp {\it et al} \cite{Kaupp1991} by a further 10 pm but this value was now 4 pm too high. 
Going one step further, Allouche {\it et al} \cite{Allouche1992}  introduced 
an $\ell$-dependent CPP into a CI calculation of the lowest electronic states of BaH using a triple zeta basis set. Although there were eleven 
valence electrons, only three were active in the CI step. The X$^2\Sigma^+$ state equilibrium bond length was still 3 pm too large, but the agreement is 
better than Kaupp {\it et al}. \cite{Kaupp1991} In addition the molecular constants for all the excited states correlating the lowest five electronic states of 
barium were calculated. They also investigated the effect of including spin-orbit coupling in the computation and demonstrated it made no difference (at the 100 fm level)  to the
 ground state bond length and altered excited states by $<$ 0.3 pm. However, the spin-orbit model adopted was fairly basic and consequently 
the spin-orbit constants were up to 60\% too large. The calculated dissociation energy, D$_e$, of 2.04 eV was significantly larger than the earlier 
value  of 1.79 eV recommended by Fuentealba\cite{Preuss1987} {\it et al} from their pseudo-potential 
plus single and double configuration interactive (CISD) model, with an equilibrium bond distance of 2.3707 $\text{\AA}$. 
We note that Fuentealba et al \cite{Preuss1987} also obtained a value of 2.1 eV for D$_e$ using a pseudo-potential 
 plus local spin density (LSD) approximation with an equilibrium bond distance of  2.4829 $\text{\AA}$. 
That dissociation energy is in excellent agreement with a recent  relativistic coupled-cluster (RCCSDT) value of 2.062 eV by Skripnokov {\it et al} \cite{Skripnikov2013}. 
This is interesting in itself as such agreement between relativistic and essentially non-relativistic calculations is rather unexpected in such a heavy hydride. Despite the close agreement between the highlighted theoretical bond energies, they are all larger than the accepted\cite{Lindgren1965} experimental value.

 Following its discovery in  the visible emission spectra\cite{Eagle1909} 
of the Group II hydrides by Eagle in 1909 
the E$^2\Pi$ $ \rightarrow$ X$^2\Sigma^+$ 0 – 0 band in 
BaH was assigned by Schaafsma \cite{Schaafsma1932} around the time
 that Watson began a series of spectroscopic studies on BaH 
recording \cite{Watson1932} the (1 - 1) and (2 - 2)  E$^2\Pi$ $ \rightarrow$  X$^2\Sigma^+$
 emission bands (Funke later \cite{Funke1933} added the 
 non-diagonal (0 – 1) and (2 - 1) bands). Watson followed that work with the 
first spectra of the B$^2\Sigma^+$ $ \rightarrow$ X$^2\Sigma^+$ transition \cite{Watson1933} in the near infra-red, which revealed 
an exceptionally large value for the spin-rotation constant 
in the upper state. As this constant was determined to be negative, 
it suggested the presence of the lower lying  A$^2\Pi$ state that was formally identified by the group \cite{Watson1935,Koontz1935} two years later. Also conjectured in these papers was the presence \cite{Watson1935} of an even lower lying excited $^2\Delta$ state, though this H$^2\Delta$ potential was finally directly observed \cite{Fabre1987} by Fabre {\it et al.} only in 1987. In 1936 Grunstr{\"o}m  \cite{Grunstrom1936} reported a new higher lying electronic state and analysed the (0 – 0) and (0 - 1) bands of this violet 
C$^2\Sigma^+$ $ \rightarrow$ X$^2\Sigma^+$ transition, a similar band having been observed in lighter Group II hydrides. 
Further investigation \cite{Funke1936} revealed that the $v^{\prime}$ = 1 level is abruptly predissociated above $N$ = 10 and it was speculated there was a lower bound state with an extended equilibrium bond length. 

A series of papers in the 1960s by Kopp and colleagues revealed absorptions in the UV \cite{Aslund1963}, identified the D$^2\Sigma^+$ state \cite{Lindgren1965} perturbing the C$^2\Sigma^+$ vibronic structure, recorded the B$^2\Sigma^+$ $ \leftarrow$ X$^2\Sigma^+$ transition in BaD \cite{Wirhed1966} and added additional vibrational levels to the A$^2\Pi$ $ \leftarrow$  X$^2\Sigma^+$ absorption spectra \cite{Guntsch1966} for both hydrogen isotopes. However, due to the highly diagonal transitions in the A$^2\Pi$, B$^2\Sigma^+$ and C$^2\Sigma^+$ $ \leftarrow$ X$^2\Sigma^+$  absorption bands, direct observations of vibrational levels were limited to $v$ = 0 - 2 for all the electronic states involved. Only the D$^2\Sigma^+$ state had a potential minimum significantly displaced from that of the ground state but has only been studied in absorption so while the D$^2\Sigma^+$ state  has been studied up to $v$ = 9, the ground state remains unexplored \cite{Bernath1993,Bernath2013} above $v$ = 3. Furthermore, the often quoted experimental value for the dissociation energy in the ground state,  D$_e$ $<$ 16350 cm$^{-1}$, is based \cite{Lindgren1965} on the analysis of predissociation in the C$^2\Sigma^+$ state and assumptions made about the nature of mechanism involved.

	%
	%
\begin{figure*}
\centering
\includegraphics[scale=0.65]{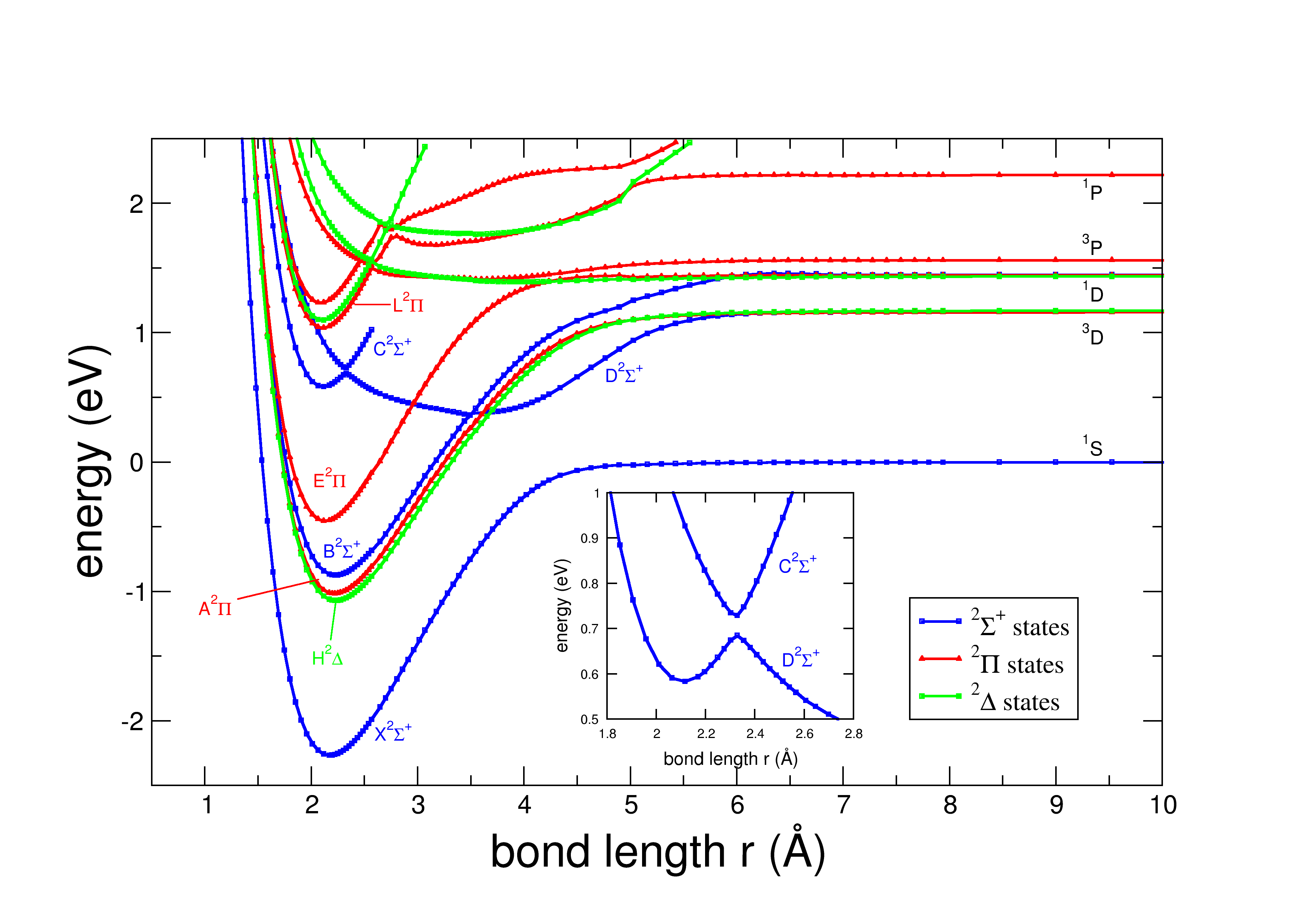}
\caption{MRCI+Q results for the lowest 5 states of each symmetry ($^2\Sigma^+$, $^2\Pi$ and $^2\Delta$) as a function of 
			bond length.  The calculations have been performed with an aug-cc-pV6Z basis set and an ECP with the MOLPRO suite.  
			The potentials shown were used to  extract 
			 values for both the equilibrium bond distance $r_e$ and dissociation energy D$_e$.  
			 For the ground X$^2\Sigma^+$ state, the calculated equilibrium bond length  is $r_e$=2.18 $\text{\AA}$ while the dissociation energy is D$_e$ = 2.2687 eV. The electronic states of the Ba atom fragments are also labelled (only H atoms in the lowest 1$^2$S state are produced at these energies).  Inset: details of the avoided crossing between C/D$^2\Sigma^+$ states.} \label{av6z}
\end{figure*}

\section{{\it Ab Initio} Calculations}

\subsection{Computational methods}
\noindent
The initial results reported here were obtained using a parallel version of the 
 MOLPRO~\cite{Werner2010} suite of {\it ab initio} quantum chemistry codes
 (release MOLPRO 2010.1).
MRCI calculations based on State-Averaged Multi-Configuration-Self-Consistent-Field 
(SA-MCSCF) wavefunctions \cite{Siegbahn1980,Werner1985,Knowles1985} were conducted to determine the potential energy curves as a function 
of internuclear distance $r$ out to  
bond separation of  $r = 40$ bohr (a$_0$). When dealing with dissociating systems, size-consistency in the calculation is a particular concern, particularly with CI methods\cite{Meissner1988} restricted to single and double excitations such as the MRCI code in MOLPRO. To estimate the missing higher excitations, the {\it a posteriori}
Davidson  correction\cite{Davidson1974} was applied to all the results (MRCI + Q).
To model the molecular orbitals a variety of correlation-consistent
Gaussian basis sets\cite{Peterson2015} were trialled; aug-cc-pV$n$Z, 
cc-pCV$n$Z-PP, cc-pwCV$n$Z-PP and aug-cc-pCV$n$Z-PP. In each calculation, the 46 core electrons of the barium atom were described 
with an effective core potential of the form ECP46MDF from the Stuttgart group\cite{Schwerdtfeger2006} leaving the outer ten 5s5p6s electrons. This allows one to test  how each basis set performs 
for the molecular constants, e.g.; the equilibrium bond length and the dissociation energy D$_e$ while keeping the active space constant.  

All the electronic structure computations are performed in the C$_{2v}$ 
Abelian point group symmetry. In the  C$_{2v}$  point group, molecular orbitals are labelled according to symmetries $a_1$,$b_1$,$b_2$ and $a_2$; when the molecular
symmetry is reduced from C$_{\infty v}$ to C$_{2v}$, the correlating relationships
are $\sigma \rightarrow a_1$, $\pi \rightarrow$ ($b_1$, $b_2$) , $\delta \rightarrow$ ($a_1$, $a_2$). 
The preferred active space consists of 11 electrons and 17 molecular orbitals $7a_1$, $5b_1$, $3b_2$, $2a_2$; i.e. a (7,5,3,2) model.
However, to describe the Rydberg character of the highly excited states of $A_1$ symmetry it was necessary sometimes to  expand the 
active space  to (9$a_1$, 5$b_1$, 3$b_2$, 2$a_2$).
The averaging process was carried out on the lowest five $^2\Sigma^+$, $^2\Pi$ and  $^2\Delta$ molecular states of this hydride.

\subsection{{\it Ab initio} potentials using an AV6Z basis}
\noindent
Typically, the basis set of choice for an MRCI + Q calculation including excited electronic states is the augmented correlation consistent type such as aug-cc-pV$n$Z to provide the diffuse Gaussians required to model the higher energy electron orbitals. In addition, the larger the value of $n$ the more faithful the model wavefunctions match the real ones. Therefore, the first computations on this hydride were performed using an aug-cc-pV6Z basis on both atoms (Fig. \ref{av6z}), extending the previous work of Lane\cite{Lane2015}.  The calculated potentials can be grouped according to three broad types:
\vskip 1mm  \noindent
(1) Bound potentials based on the interaction between a Ba$^+$ cation and H$^-$ anion e.g. the X$^2\Sigma^+$ and A$^2\Pi$ states, \vskip 0.5mm  \noindent
(2) Rydberg states with a single electron held within the potential of a BaH$^+$ molecular ion composed of Ba$^{2+}$ and H$^-$ ions e.g. C$^2\Sigma^+$ and \vskip 0.5mm  \noindent
(3) extended bond length states formed by the avoided crossing of largely repulsive valence curves and attractive ionic potentials e.g. D$^2\Sigma^+$. 
\vskip 1mm 	\noindent
Only doublet states were investigated as the lowest quartet states\cite{Allouche1992} are all repulsive. The historical spectroscopic assignments were used for the electronic states in Fig. \ref{av6z} although it is clear that the adiabatic 3$^2\Sigma^+$ potential is reported as two separate states experimentally (C$^2\Sigma^+$ and D$^2\Sigma^+$). The calculated energy gaps at the majority of the avoided crossings in BaH are tiny (the case of the 2$^2\Sigma^+$-X$^2\Sigma^+$ adiabats being a noteworthy exception) and consistent with experiments where, for example, the Rydberg C$^2\Sigma^+$ vibrational progression\cite{Lindgren1965} is maintained for energies well above that of the C/D$^2\Sigma^+$ states' closest approach (inset, Fig. \ref{av6z}). These AV6Z $\Lambda$S$\Sigma$ potentials are consistent with previous work by Allouche\cite{Allouche1992} {\it et al}.  

For the ground X$^2\Sigma^+$ state the calculated
value of $r_e$ = 2.18~$\text{\AA}$ is in close agreement to the  value 
of $r_e$ = 2.16 $\text{\AA}$ obtained\cite{Lane2015} using a barium triple zeta basis set (cc-pV5Z basis on the hydrogen), a 56-electron ECP and an $\ell$ independent 
core polarisation potential (CPP). However, the accepted experimental value\cite{Bernath2013} 
for $r_e$  is 2.2319 $\text{\AA}$. The AV6Z basis set also predicts a dissociation energy D$_e$ of 2.2687 eV (18293.08 cm$^{-1}$) while the experimental value reported by Kopp {\it et al} \cite{Lindgren1965} is considerably smaller,
$<$ 2.0271~eV (16350 cm$^{-1}$).  Both the discrepancies in the equilibrium bond length and the dissociation energy are rather large for an MRCI+Q calculation with such a substantial barium basis set. Therefore, the investigation was widened to include  
a variety of other basis sets in an attempt to improve the agreement between the {\it ab initio} results for the 
ground state for this hydride and the spectroscopic constants.

\subsection{Augmented basis sets}
\noindent
An even tempered augmentation can be used to extend both the cc-pCVQZ-PP and the cc-pCV5Z-PP basis\cite{Peterson2015} sets. Each basis set is augmented with a single diffuse Gaussian function, the augmented exponent $\alpha^{\prime}$ given by the expression,
\begin{equation}
\alpha^{\prime} (\ell) = \frac{\alpha^2(\ell)}{\beta(\ell)}
\end{equation}
where $\alpha(\ell)$ and $\beta(\ell)$  are respectively the last and second last exponent of 
the appropriate basis set. Calculations are performed in the MRCI+ Q approximation  
for the even tempered augmented sets (the additional functions are presented in 
Table \ref{aug-tempered}) and explicitly optimized augmented cc-pCV$n$Z-PP ($n = 4$ and 5) sets\cite{Peterson2015} (these explicitly optimized functions 
are presented in Table \ref{aug-optimized}). A large augmented basis set is chosen because in quantum chemistry calculations they  
are known to recover typically $\sim$ 99 \% or more of the electron correlation energy.
	%
	%
\begin{table}
\centering
\caption{Augmented functions to the cc-pCV$n$Z-PP basis set \cite{Peterson2015} for barium.
	    The even tempered augmentation proceedure is used, where $\alpha(\ell)$ and $\beta(\ell)$ 
	    are respectively the last and second last exponent in appropriate basis set.}
	    
\vskip 2mm
\label{aug-tempered}
\begin{tabular}{ccccc}
\hline\hline \noalign{\vskip 1.2mm}
	cc-pCVQZ-PP			 		& Barium &	\\
\noalign{\vskip 1mm}
Angular momentum function       		&&    Exponent$^a$  \\
$s$                                			 &&  0.32620	\\	
$p$                               			 &&  0.10943  \\
$d$                               			 & &  0.18940    \\
$f$                               			& &  0.43550    \\
$g$                             			& &  0.47145  \\
\\
	cc-pCV5Z-PP				 & Barium &	\\
\noalign{\vskip 1mm}
Angular momentum function       		&&    Exponent$^a$  \\
$s$                                			 &&  0.18570	\\	
$p$                               			 &&  0.11590  \\
$d$                               			 & &  0.29510    \\
$f$                               			& &  0.38060    \\
$g$                             			& &  0.46480  \\
$h$						&&   0.62250 \\
\hline\hline
\end{tabular}
\begin{flushleft}
$^a$Estimated using $\alpha^{\prime} (\ell) = \frac{\alpha^2(\ell)}{\beta(\ell)}$
\end{flushleft}
\end{table}
\begin{table}
\centering
\caption{Augmented functions to the cc-pCV$n$Z-PP basis set for barium.
	    For the augmentation proceedure the exponents are optimised\cite{Peterson2015} on the energy of the 
              ground state of the BaH molecule.}
\vskip 2mm
\label{aug-optimized}
\begin{tabular}{ccccc}
\hline\hline \noalign{\vskip 1.2mm}
		cc-pCVQZ-PP		 	& Barium &	\\
\noalign{\vskip 1mm}
Angular momentum function       		&&    Exponent \\
$s$                                			 &&  0.0073	\\	
$p$                               			 &&  0.0060  \\
$d$                               			 & & 0.0114    \\
$f$                               			& &  0.0373    \\
$g$                             			& &  0.0492  \\
\\
		cc-pCV5Z-PP			&Barium	&	\\
\noalign{\vskip 1mm}
Angular momentum function		&& Exponent   \\
$s$                             			& &0.0071  \\
$p$                             			& &0.0058 \\
$d$                            			& &0.0089\\
$f$                             			& & 0.0326\\
$g$						& & 0.0393\\
$h$						& & 0.0515\\
\hline\hline
\end{tabular} 
\end{table}
	%
	%
	%
\begin{table*}
\centering
\caption{Spectroscopic constants for the X$^2\Sigma^+$ ground state of the BaH molecule:  
	     the equilibrium bond distance $r_e$($\text{\AA}$) and
	     the dissociation energies D$_e$ in eV and cm$^{-1}$.  
	     Comparison of the present  MRCI + Q results with a variety of other reported values using correlation - consistent  
	     basis sets \cite{Peterson2015} (cc-pV$n$Z-PP,  aug-cc-pV$n$Z-PP,  cc-pCV$n$Z-PP, cc-pwCV$n$Z-PP, 
	    and aug-cc-pCV$n$Z-PP), both with CI and CCSDT methods (fc, is frozen core, all, is all-electron) and other 
               previous theoretical work. The best experimental values are also included.}
              
\label{bah-results}
\begin{tabular}{lcccccccccccc}
\noalign{\vskip 1mm}
\hline	\hline \noalign{\vskip 1mm} 
Basis		 		&&& Method					&&&$r_e(\text{\AA})$		&&&D$_e$ (eV) 		&&&D$_e$ (cm$^{-1}$) 	\\
\hline
\noalign{\vskip 2mm}
cc-pVQZ-PP			&&&MRCI+Q					&&&2.2039			&&&2.2229			&&&17929.31		\\
cc-pV5Z-PP			&&&MRCI+Q					&&&2.2329			&&&2.5961			&&&20939.44		\\
CBS					&&&MRCI+Q					&&&2.2376			&&&2.6580			&&&21438.71		\\
\\
aug-cc-pV6Z			&&&MRCI+Q					&&&2.1800			&&&2.2687			&&&18293.08		\\
\\
cc-pCVQZ-PP			&&&MRCI+Q					&&&2.2367			&&&2.0543			&&&16569.43		\\
cc-pCV5Z-PP			&&&MRCI+Q					&&&2.2151			&&&2.5829			&&&20832.97		\\
CBS					&&&MRCI+Q			          	 	&&&2.2118			&&&2.6715			&&&21547.60		\\
\\
cc-pwCVQZ-PP		&&&MRCI+Q					&&&2.2367			&&&2.0587			&&&16604.92		\\
cc-pwCV5Z-PP		&&&MRCI+Q					&&&2.2118			&&&2.5954			&&&20933.80		\\
CBS					&&&MRCI+Q					&&&2.2080			&&&2.6854			&&&21659.71		\\
\\
aug-pCVQZ-PP		&&&MRCI+Q$^a$				&&&2.2354			&&&2.0814			&&&16787.75		\\
aug-pCV5Z-PP 		&&&MRCI+Q$^a$				&&&2.2332			&&&2.0871			&&&16833.78		\\
CBS					&&&MRCI+Q$^a$				&&&2.2322			&&&2.0946			&&&16895.12		\\
\\
aug-pCVQZ-PP		&&&MRCI+Q$^b$				&&&2.2273			&&&2.1157			&&&17064.66		\\
aug-pCV5Z-PP		&&&MRCI+Q$^b$				&&&2.2204			&&&2.2384			&&&18054.33		\\
CBS					&&&MRCI+Q$^b$				&&&2.2187			&&&2.2278 			&&&17968.83		\\	
\\
cc-pVQZ-PP			&&&CCSD(T), fc$^c$				&&&2.3302			&&&2.0311 			&&&16382.30		\\
cc-pV5Z-PP			&&&CCSD(T), fc$^c$				&&&2.3304			&&&2.0563			&&&16585.56		\\
\\
cc-pCVQZ-PP			&&&CCSD(T), fc$^c$				&&&2.3266			&&&2.0494			&&&16529.91		\\
cc-pCV5Z-PP			&&&CCSD(T), fc$^c$				&&&2.3280			&&&2.0632  			&&&16641.21		\\
\\
cc-pwCVQZ-PP		&&&CCSD(T), fc$^c$				&&&2.3279			&&&2.0498 			&&&16533.13		\\
cc-pwCV5Z-PP		&&&CCSD(T), fc$^c$				&&&2.3282			&&&2.0632  			&&&16641.21		\\
\\
cc-pCVQZ-PP			&&&CCSD(T), all$^c$				&&&2.2378			&&&2.0298 			&&&16371.82		\\
cc-pCV5Z-PP			&&&CCSD(T), all$^c$				&&&2.2316			&&&2.0723			&&&16714.61 		\\
\\
cc-pwCVQZ-PP		&&&CCSD(T), all$^c$				&&&2.2379			&&&2.0355			&&&16417.79		\\
cc-pwCV5Z-PP		&&&CCSD(T), all$^c$				&&&2.2327			&&&2.0736			&&&16725.10		\\
\\
aug-cc-pVQZ			&&&RCCSD(T)$^d$				&&&2.2400			&&&2.0886			&&&16846.08		\\
aug-cc-pV5Z			&&&MRCI+Q+SOC$^e$			&&&2.2417			&&&2.0860			&&&16825.11		\\
\\
Triple-zeta			&&&CIPSI (ECP+CPP)$^f$		&&&2.2620			&&&1.9700			&&&15889.49		\\
					&&&Pseudo-potential + LSD$^g$	&&&2.4280			&&&2.1000			&&&16938.03		\\
					&&&Pseudo-potential + CISD$^g$	&&&2.3707			&&&1.7900			&&&14437.66		\\
\\
Experiment			&&&						&&&2.2319$^h$		&&&$<$2.0271$^i$	&&&$<$16350$^j$		\\
\\
cc-pCVQZ-PP			&&& MRCI+Q  (3e)$^k$			&&&2.23670			&&&2.0543			&&&16569.43		\\
cc-pCVQZ-PP			&&& MRCI+Q  (5e)$^l$			&&&2.23725			&&&2.1359			&&&17227.59		\\
\noalign{\vskip 1mm}
\hline\hline
\end{tabular}
\begin{flushleft}
$^a$MRCI+Q, even tempered basis set augmentation functions, present work.\\
$^b$MRCI+Q, explicitly energy optimized augmentation functions, present work.\\
$^c$CCSD(T), fc  and CCSD(T), all : private communication from Peterson (2015).\\
$^d$RCCSD(T), Skripnikov and co-workers \cite{Skripnikov2013}.\\
$^e$MRCI+Q + SOC, Gao {\it et al} \cite{Gao2014}.\\
$^f$CIPSI Allouche \cite{Allouche1992}.\\
$^g$Pseudo-potential + local spin density (LSD) \cite{Preuss1987}, Pseudo-potential + CISD \cite{Preuss1987}.\\
$^h$Experiment, Ram and Bernath \cite{Bernath2013}.
$^i$Experiment, Kopp {\it et al} \cite{Lindgren1965}.
$^j$Assuming Ba($^3$D$_3$) limit.\\
$^k$MRCI+Q, active 3-electron model, present work.
$^l$MRCI+Q, active 5-electron model, present work. 

\end{flushleft}
\end{table*}
 
\subsection{CCSD(T) versus MRCI+Q}
\noindent
In the search for an ideal basis set, numerous calculations 
with the MOLPRO suite were performed for the ground state of this hydride. 
 Table \ref{bah-results} presents results for the different basis sets used   
in the MRCI+Q approximation, both for the equilibrium bond distance $r_e$ and dissociation energy D$_e$, and compares them to  
 CCSD(T) calculations\cite{Peterson2015} on BaH.  CCSD(T) is currently regarded as the most reliable theoretical method for information on ground state potentials. In the MRCI+Q calculations, the CASSCF step is performed over the lowest three $^2A_1$, $^2B_1$, and $^2A_2$ states while only the lowest $^2A_1$ state appears in the MRCI code. It appears that 
the even tempered augmented basis set aug-cc-pCV5Z-PP performs best at MRCI+Q level for the ground state, matching within 0.1pm the experimental $r_e$ value. 
The explicitly optimized augmented basis set aug-pCV5Z-PP gives comparable equilibrium bond distances. In contrast to calculations using CCSD(T), the cc-pCV$n$Z-PP basis sets pull the equilibrium bond lengths to much shorter values when using the MRCI + Q method. Augmenting the basis set with diffuse orbitals helps to counteract this effect.
From Table \ref{bah-results} the CCSD(T), all electron, work \cite{Peterson2015} provides results (with either the cc-pCV5Z-PP or cc-pwCV5Z-PP basis) 
comparable to the present MRCI+Q results with the even tempered augmented basis set. The MRCI + Q calculations are also in excellent agreement 
with the previous RCCSDT \cite{Skripnikov2013} and MRCI+Q + including spin-orbital coupling (SOC) \cite{Gao2014} values for the dissociation energy.  In addition,
this optimal basis set can also be used to carry out calculations of excited states, such as A$^2\Pi$, a significant advantage of the MRCI method over CCSD(T).

	%
	%
\begin{figure}
\centering
\includegraphics[scale=0.11]{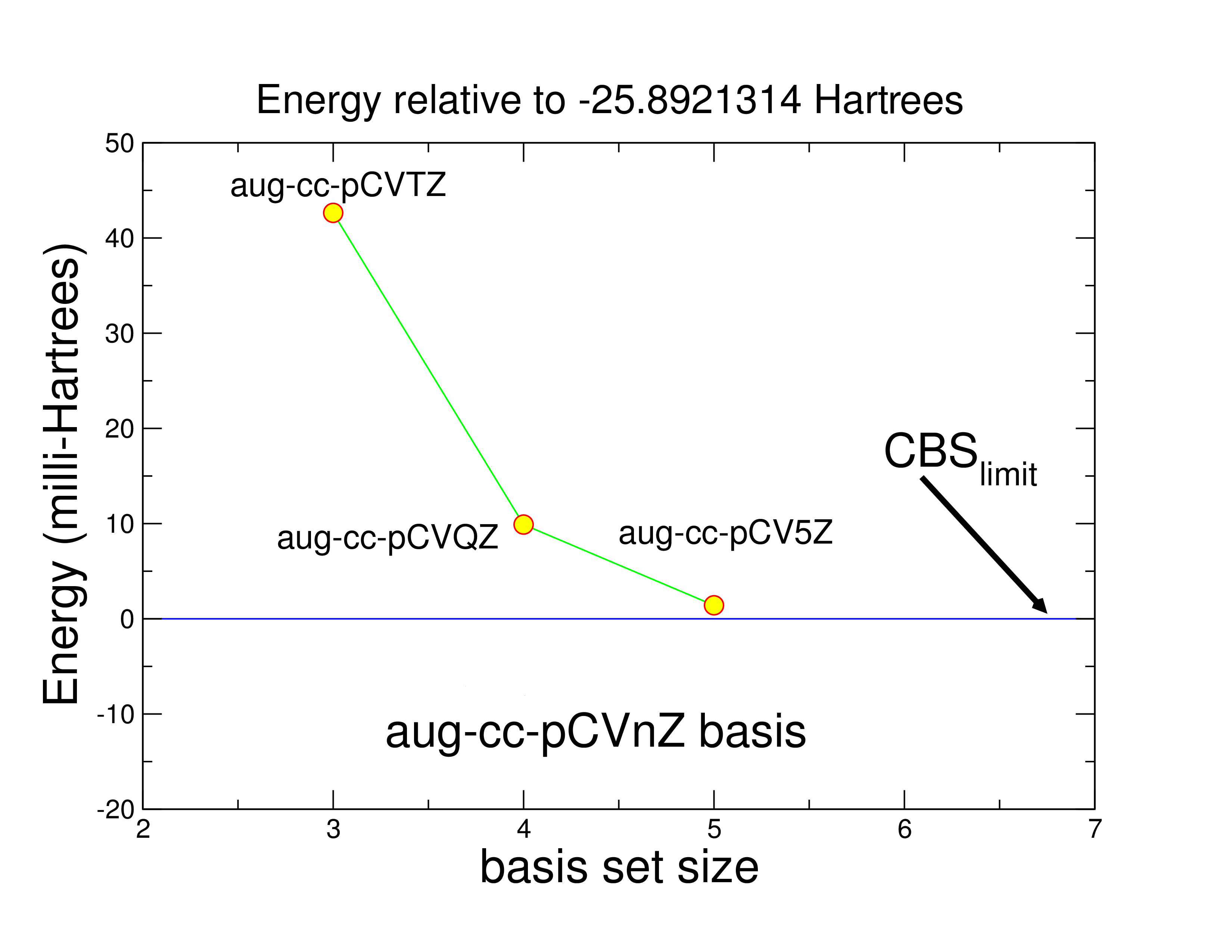}
\caption{BaH minimum energy in X $^2\Sigma^+$ state, the approach to the CBS limit  using an active 3-electron  model in the MRCI+Q approximation.  Even tempered aug-cc-pCV$n$Z-PP basis sets are used in the calculations.} \label{cbs}
\end{figure}
\subsection{Complete Basis Set Limit}
\noindent
To remove the uncertainty that any close agreement with experiment is the result of computational artefacts, it is best to extrapolate the energies E$_n$ calculated with the V$n$Z basis sets to the limit of an infinite number of Gaussian functions. 
For the various  basis sets shown in Table \ref{bah-results}
the potential  energy of the ground state 
via the complete basis set limit (CBS) is found using the formulae derived by Jensen \cite{Jensen2005} with modifications by Karton and Martin \cite{Karton2006}
\begin{equation}
E_{CBS}  = E_n   + \frac{( E_n   -   E_{n-1} )}{\frac{n}{n+1} exp[9(\sqrt{n} - \sqrt{n-1}) - 1]}
\end{equation}
where here $n$ = 5. This expression is used to determine all the CBS potential energy curves in this study. Fig \ref{cbs} shows the orderly approach to the CBS energy limit for the even tempered augmentation of the cc-pCV$n$Z-PP basis\cite{Peterson2015} set. While this extrapolation is usually adopted for static-correlation only, the small dynamic correlation found in the BaH calculations prompted its use here. The close agreement with the experimental r$_{e}$ value found in the CBS potential supports this choice. 

In Table \ref{bah-results} the results for the equilibrium bond distance $r_e$ and dissociation energy 
D$_e$ from these CBS limit, MRCI+Q calculations are compared with the coupled cluster appromixation 
CCSD(T) \cite{Cizek1966,Bartlett1982,Bartlett1993,Peterson2015}, in its various implementations 
and with other theoretical methods \cite{Allouche1992,Skripnikov2013,Andrews2004}. The density functional theory (DFT) work of Wang and Andrews \cite{Andrews2004} using a 6-311++G ($3df,3pd$) basis yielded a value of 2.2520~$\text{\AA}$ 
and second order M{\o}ller Plesset perturbation theory gave 2.2440~$\text{\AA}$ \cite{Andrews2004} respectively for the equilibrium bond distance. Gao {\it et al} \cite{Gao2014} have conducted calculations for the ground X$^2\Sigma^+$ and the first excited state A$^2\Pi$ 
of this hydride at the MRCI+Q+SOC level.  
Their {\it ab initio}  results for the spectroscopic constants $r_e$ (2.2417~$\text{\AA}$) and D$_e$ (2.0860~eV) of BaH (X$^2\Sigma^+$)
ground state are similar to those obtained from the relativistic - coupled-cluster results\cite{Skripnikov2013} (RCCSDT),
respectively 2.24~$\text{\AA}$ and 2.0840~eV, and in reasonable agreement with the available experimental results \cite{NIST2005}. The other {\it ab initio} results of great accuracy are those using CCSD(T) where the cc-pCV5Z-PP (all electron) result is within 0.02 pm of the experimental value \cite{Bernath2013} but if extrapolated to the CBS limit this agreement will be somewhat poorer. Using an aug-cc-pCV$n$Z/CBS MRCI+Q calculation the {\it ab initio} equilibrium bond distance here is within 0.031 pm of the best spectroscopic value. 

\section{Modelling the ground state}

\subsection{Spectroscopic constants of the ground state}
\noindent
An important requirement is ensuring that the final potential behaves correctly at all interatomic distances, particularly at long-range where dispersion forces are dominant\cite{Mitroy2010}.  In atomic units, the asymptotic atom-atom potential
has the form \cite{Dalgarno2000}
\begin{equation}
V (r) \simeq V_{\infty} - \left( \frac{C_6}{r^6} + \frac{C_8}{r^8}  \ldots \right) 
\end{equation}
where $C_6$, $C_8$ etc. are the
polarizabilities with $V_{\infty}$ the asymptotic limit (atomic products). $r$ is the atom-atom internuclear separation. The dynamic polarization of the ground state, which allows the computation of the leading $C_6$ term (148 a.u.), has been calculated\cite{Derevianko2010} by Derevianko {\it et al}. To calculate the following $C_8$ term, the formula derived by Tang\cite{Tang1969} requires the quadrupole polarizabilities but there are two theoretical values\cite{Patil2000,Porsev2006}  for Ba that produce $C_8$ values that differ by more than 60$\%$. 
{The smaller value, 11242 a.u., is consistent with the corresponding term calculated\cite{Patil1997} for the neighbouring hydride CsH (11710 a.u.), and is therefore the value recommended at present. 
There is no theoretical  or experimental value at present for $C_{10}$ in BaH so the expansion was truncated at $C_8$: the corresponding calculated value\cite{Patil1997} in CsH is $1.194 \times 10^6$ a.u.} 

	%
\begin{figure}
\centering
\includegraphics[scale=0.35]{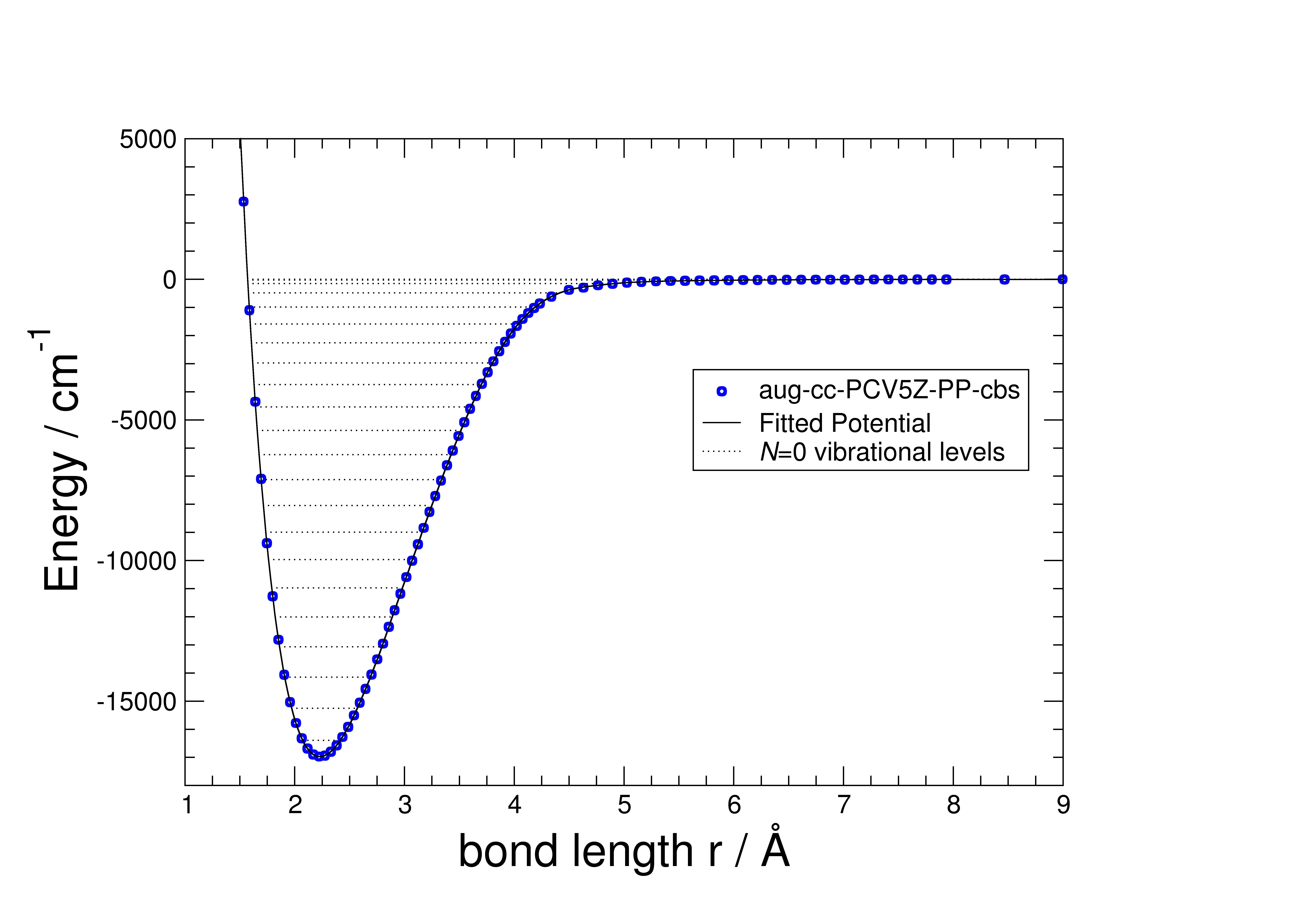}
\caption{Analysis of {\it ab initio} X$^2\Sigma^+$ ground state of BaH using \texttt{LEVEL}\cite{LeRoy2014}. The {\it ab initio} data was taken with the aug-cc-pCV$n$Z basis sets ($n$ = 4 and 5) and extrapoliated to the CBS limit. The vibrational levels found in the fit are also presented.} \label{Levels}
\end{figure}
	%
	%
	%

	%
	%
\begin{figure}
\centering
\includegraphics[scale=0.106]{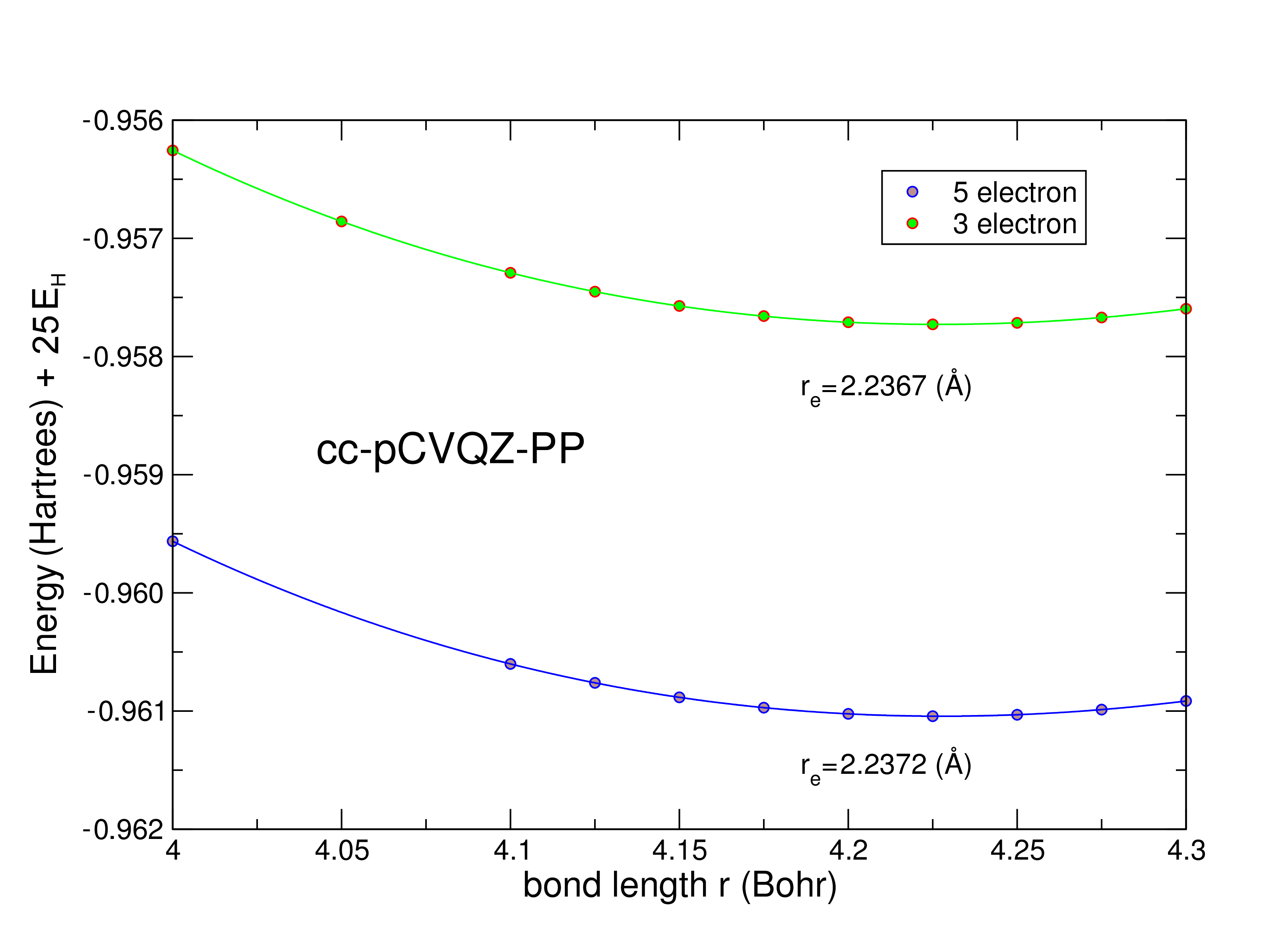}
\caption{Comparison of the ground state energy for the BaH molecule in the region of the equilibrium bond distance 
			      using both an active 3-electron and 5-electron model
			   in the MRCI+Q approximation.  The cc-pCVQ-PP basis set\cite{Peterson2015} is used for both calculations. The absolute energies are adjusted by 25 E$_H$.} \label{3e-5e}
\end{figure}
\begin{table*}
\centering
\caption{Vibrational term energies in the X$^2\Sigma^+$ ground state of the BaH molecule.  
	   Comparison of the present  MRCI+Q results with the aug-cc-pCV$n$Z-PP even tempered basis set taken to the CBS limit and experimental data. All energies in cm$^{-1}$.}
              
\label{bahX-results}
\begin{tabular}{rccrccrccrccrccr}
\noalign{\vskip 1mm}
\hline	\hline \noalign{\vskip 1mm} 
$v^{ \prime \prime}$		 		&&& MRCI 		 &&&Relative value$^{a}$
&&&Experimental\cite{Bernath2013}  &&& RKR$^{b}$ &&& $\Delta$E RKR \\	
\hline
\noalign{\vskip 2mm}
0		&&&		581.3491		&&&	0				&&&	0			&&&	580.53		&&&	0.8191		\\
1		&&&		1722.2738		&&&	1140.9247		&&&	1139.2896	&&&	1719.84		&&&	2.4338		\\
2		&&&		2832.9237		&&&	2251.5746		&&&	2249.6062	&&&	2830.15		&&&	2.7737		\\
3		&&&		3915.9012		&&&	3334.5521		&&&	3331.1192	&&&	3911.46		&&&	4.4412		\\
4		&&&		4981.7132		&&&	4400.3641		&&&				&&&	4963.77		&&&	17.9432		\\
5		&&&		6024.9686		&&&	5443.6195		&&&				&&&	5987.08		&&&	37.8886		\\
6		&&&		7029.1845		&&&	6447.8354		&&&				&&&	6981.39		&&&	47.7945		\\
7		&&&		8006.0703		&&&	7424.7212		&&&				&&&	7946.70		&&&	59.3703		\\
8		&&&		8957.6993		&&&	8376.3502		&&&				&&&	8883.01		&&&	74.6893		\\
9		&&&		9878.6102		&&&	9297.2611		&&&				&&&	9790.32		&&&	88.2902		\\
10		&&&		10772.2217		&&&	10190.8726		&&&				&&&	10668.63	&&&	103.5917	\\
11		&&&		11633.1528		&&&	11051.8037		&&&				&&&				&&&				\\
12		&&&		12461.4104		&&&	11880.0613		&&&				&&&				&&&				\\
13		&&&		13257.3637		&&&	12676.0146		&&&				&&&				&&&				\\
14		&&&		14016.6614		&&&	13435.3123		&&&				&&&				&&&				\\
15		&&&		14729.2373		&&&	14147.8882		&&&				&&&				&&&				\\
16		&&&		15386.3161		&&&	14804.9670		&&&				&&&				&&&				\\
17		&&&		15966.4205		&&&	15385.0714		&&&				&&&				&&&				\\
18		&&&		16436.4095		&&&	15855.0604		&&&				&&&				&&&				\\
19		&&&		16749.4237		&&&	16168.0746		&&&				&&&				&&&				\\
20		&&&		16882.9819		&&&	16301.6328		&&&				&&&				&&&				\\
21		&&&		16894.9472		&&&	16313.5981		&&&				&&&				&&&				\\
\noalign{\vskip 1mm}
\hline \hline
\end{tabular}
\begin{flushleft}

$^a$E(MRCI) - E(experimental) = $\Delta$E ($v^{\prime}$ - $v^{\prime \prime}$) cm$^{-1}$: (1 - 0) = +1.64; (2 - 1) = +0.33; (3 - 2) = +1.46. \\
$^b$Based on experimental data prior\cite{Ramanaish1982} to 1982. \\

\end{flushleft}
\end{table*}

	%
	%
\begin{figure}
\centering
\includegraphics[scale=0.35]{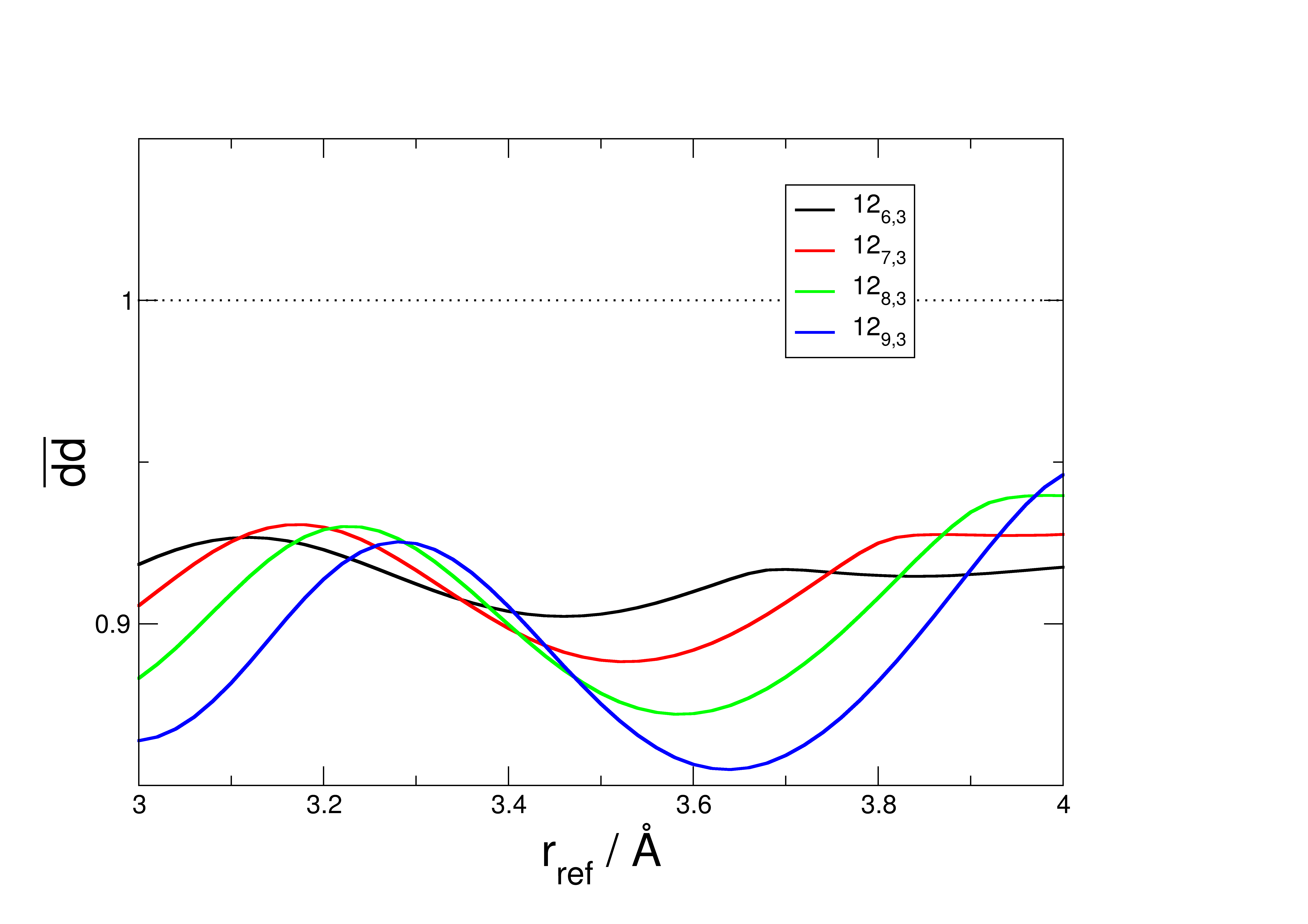}
\caption{Tuning r$_{ref}$ to improve the fitting of the {\it ab initio} X$^2\Sigma^+$ ground state of BaH using \texttt{betaFIT}\cite{LeRoy2013}.} \label{Rref}
\end{figure}

The \texttt{LEVEL}\cite{LeRoy2014} program (Version 8.2) was used to calculate the rovibrational levels of the resulting \textit{ab initio} X$^2\Sigma^+$ aug-cc-pCV$n$Z/CBS potential. A smooth potential was produced through interpolation of the {\it ab initio} points with a 4-point piecewise polynomial (\texttt{NUSE}=4, \texttt{IR2}=1), and extrapolation past $r = 15.88$ $\text{\AA}$ using a fixed $C_6$ = 7.110512 $\times 10^5$ cm$^{-1}$ $\text{\AA}^{-6}$ (no $C_8$ or $C_{10}$) (\texttt{ILR}=2).
Based on this potential the equilibrium bond length and dissociation energy were determined to be 2.2322 {$\text{\AA}$} and 16895 cm$^{-1}$. Twenty-two vibrational levels were predicted for the system, with $v = 21$ lying just 0.2 cm$^{-1}$ below the dissociation limit. Comparison of the relative energies of the first four vibrational levels based on this potential with experimental values\cite{Bernath2013} show good agreement, all within 1.7 cm$^{-1}$ (Table \ref{bahX-results}). There is a significant jump in the error with respect to the earlier RKR levels at $v$ = 4, a result of the lack of experimental data for the higher vibrational levels and consequently the present results are probably more reliable for those missing vibrational levels.
\subsection{3 electron versus 5 electron MRCI+Q}
\noindent
Though the calculated bond length is only $<$0.031 pm away from the experimental value, another effort was made to explore possible improvements to the accuracy of the {\it ab initio} potential. In order to quantify the effect of including additional dynamic electron correlation, further MRCI+Q calculations were performed with the cc-pCVQZ-PP basis set \cite{Peterson2015} using both 3- and 5-electrons active at the CI stage. 
From the {\it ab initio} work shown in Fig.~\ref{3e-5e} it is found that the active 5-electron  
model  lowers the absolute (total) energy of the ground state minimum for this hydride by just 3~milli-Hartrees (mE$_h$).  
The subsequent change in the equilibrium bond distance  $r_e$  was also tiny, an increase of 0.055~pm. 	
These very small changes in the total energy and equilibrium distance 
subsequently justifies the use of the 3-electron model for both the ground 
and excited states considered as opposed to the more computationally intensive active 5-electron model. 

\subsection{Including spectroscopic data}
\noindent
There is an extensive selection of pair-potentials in the literature to describe the $r$ dependence of the interaction energy in a diatomic molecule (see the supplementary information from Xie {\it et al} for a comprehensive list\cite{Xie2014} with over a hundred examples). Two of the most successful functions are the Tang-Toennies\cite{Tang1984} potential (and its variants) and the Morse-Long-Range\cite{LeRoy2006a} (MLR) function. The {\it ab initio} data here was fitted to a MLR potential using \texttt{betaFIT} (version 2.1)\cite{LeRoy2013} in order to find an analytical function to express the potential. Prior to fitting, the potential was shifted by 0.03 pm so that the {\it ab initio} potential minimum was aligned with the experimental r$_e$ value to an accuracy better than 0.01~pm. This simplifies the inclusion of experimental data at a later stage. The MLR potential is described by
\begin{equation}
V_{\text{MLR}} (r) \equiv \mathfrak{D}_e \left(1 - \frac{u(r)}{u(r_e)} e^{-\beta(r) y_p^{r_e} (r)}\right)^2
\end{equation}
where $\mathfrak{D}_e$ is the dissociation energy and $u(r_e)$ is value of $u(r )$ at the equilibrium bond distance. 

Both V$_\text{MLR}$ and $\beta (r)$ are expanded in terms of two internuclear distances r$_e$ and r$_{ref}$, according to the radial variable
\begin{equation}
y_n^{r_{i}} (r) = \frac{r^n - r^n_{i}}{r^n + r^n_{i}}
\end{equation}
  where r$_i$ = r$_e$ or r$_{ref}$ and the latter is usually greater than or equal to r$_e$. 
  {As the fit is viable for a large range of r$_{ref}$, it is worth tuning this value to improve the fit (Fig. \ref{Rref}). MLR functions producing local minima were selected as starting points for later fitting to spectral data.}
  Typically, two different values of $n$ are required for an acceptable fit (in the present work, $n = 8$ and 3) in the $\beta (r)$ expansion. The first issue is what to set as the error in the ab initio points. Since the vibrational frequencies are calculated to within an average of 2 cm$^{-1}$ the error in each {\it ab initio} point was initially set rather arbitrarily as 10 cm$^{-1}$ ($5 \times 10^{-5}$ E$_h$). However, Dattani and Le Roy\cite{Dattani2015a} have shown that even in the 6e Li$_2$ molecule a deviation in D$_e$ of up to 68 cm$^{-1}$ still persists in the calculated b$^{1}\Pi$ state. Therefore, an increased error of 20 cm$^{-1}$ was adopted as a compromise value.  The final analytical potential is reported as MLR$_{p,q}^{r_{ref}}$ ($N_{\beta}$), where $p$ and $q$ are the two values of $n$ required and consists of $N_{\beta}$ expansion coefficients. A similar\cite{Henderson2013} potential (replacing an earlier version\cite{Shayesteh2007} that used only one expansion co-efficient $p$) has recently been used  to describe the ground state of MgH, the only hydride other than H$_2$ with an experimental dissociation energy determined to sub-wavenumber accuracy. The fitted $\mathfrak{D}_e$ is just over {4~cm$^{-1}$ lower than the result from the {\it ab initio} potential (Table \ref{DPotParam}), as it was allowed to vary slightly during the fitting process.}

The parameters fitted to {MLR$_{8,3}^{3.34}$} ($12$) from the {\it ab initio} potential were then transferred to \texttt{DPotFit}\cite{LeRoy2009,LeRoy2013a} (version 2.0) and combined with B$^2 \Sigma^+ \rightarrow$ X$^2 \Sigma^+$ emission\cite{Watson1933,Koontz1935,Wirhed1966} and  X$^2 \Sigma^+$ infrared data\cite{Bernath1993}. {The earlier emission data was quoted to 0.01 cm$^{-1}$ while the stated resolution of the infrared data was 0.005 cm$^{-1}$.}

{While r\(_{ref} = 3.34 \)~\AA\ does not correspond to a local minimum in Fig. \ref{Rref}, those MLRs obtained from the local minima in \texttt{betaFIT} were not found to be as successful. $p=8$, $q=3$ were selected as the best parameters after screening other combinations.}
As \texttt{DPotFit} utilizes a dimensionless root-mean-square-deviation ($\overline{dd}$) as its means of determining how closely a model matches the observed data, it is crucial that a reasonable assessment of the uncertainty in the observable data is made.
\begin{equation}
\overline{dd} = \sqrt{\frac{1}{N_\text{data}}\sum_{i=1}^{N_\text{data}} \left(\frac{E_\text{calc} (i) - E_\text{obs} (i)}{u_\text{obs} (i)}\right)^2 }
\end{equation}
Thus, as the uncertainty in the {\it ab initio} E$_\text{obs}$ was set to {20} cm$^{-1}$, deviations from the fitted MLR do not lead to a large increase in the $\overline{dd}$. At the chosen r$_{ref}$ the $\overline{dd}$ between the {MLR$_{8,3}^{3.34}$}($12$) potential fitted by \texttt{betaFIT} and the {\it ab initio} points was only {0.4570} representing a close fit within the selected uncertainty. By contrast the original emission data was quoted to much higher accuracy, so 
any deviations between the model and the spectral measurements can lead to a significant increase (left hand data, Table \ref{DPotParam}) in $\overline{dd}$. {Due to the arbitrary nature of the uncertainty assignment in the {\it ab initio} data, the associated $\overline{dd}$ should not be compared directly to the $\overline{dd}$ in fits to experimental data.} 
\begin{table*}
\centering
\caption{Summary of the spectral data used when fitting with \texttt{DPotFit}. A total of 1769 B\(^2\Sigma\)-X\(^2\Sigma\) transition and 409 X\(^2\Sigma^+\) IR emission lines were incorporated.}
\label{tab-BXspecinfo}
{
\begin{tabular}{lcccccccccccc}
\noalign{\vskip 1mm}
\hline \hline \noalign{\vskip 1.2mm}
Year					&&	Transition						&&	Isotope								&&	\(v'-v'' \left(J_{max}\right)\)	&&	Lines	&& Unc / cm\(^{-1}\)	\\
\hline \noalign{\vskip 1mm}
1933\cite{Watson1933}	&&	B\(^2\Sigma^+\) $\rightarrow$ X\(^2\Sigma^+\)	&&	\(^{138}\)Ba\(^1\)H					&&	0-0 (37.5)						&&	112			&&	0.01	\\
\noalign{\vskip 1mm}
1935\cite{Koontz1935}	&&	B\(^2\Sigma^+\) $\rightarrow$ X\(^2\Sigma^+\)	&&	\(^{138}\)Ba\(^1\)H					&&	1-0 (41.5)						&&	142			&&	0.01	\\
						&&								&&										&&	1-1 (31.5)						&&	106			&&	0.01	\\
						&&								&&										&&	2-1 (37.5)						&&	126			&&	0.01	\\
\noalign{\vskip 1.2mm}
1966\cite{Wirhed1966}	&&	B\(^2\Sigma^+\) $\leftarrow$ X\(^2\Sigma^+\)	&&	\(^{138}\)Ba\(^2\)H					&&	0-0 (52.5)						&&			197	&&	0.01	\\
						&&									&&										&&	1-0 (47.5)						&&			177	&&	0.01	\\
						&&									&&										&&	0-1 (42.5)						&&			146	&&	0.01	\\
						&&									&&										&&	1-1 (50.5)						&&			179	&&	0.01	\\
						&&									&&										&&	2-1 (46.5)						&&			162	&&	0.01	\\
						&&									&&										&&	1-2 (33.5)						&&			94	&&	0.01	\\
						&&									&&										&&	2-2 (44.5)						&&			142	&&	0.01	\\
						&&									&&										&&	3-2 (39.5)						&&			125	&&	0.01	\\
						&&									&&										&&	3-3 (29.5)						&&			 61	&&	0.01	\\
\noalign{\vskip 1mm}
1993\cite{Bernath1993}	&&	IR	&&	\(^{138}\)Ba\(^1\)H					&&	1-0 (28.5)						&&			94	&&	0.005	\\
						&&								&&										&&	2-1 (20.5)						&&			 80	&&	0.005	\\
						&&								&&										&&	3-2 (29.5)						&&			 77	&&	0.005	\\
\noalign{\vskip 0.5mm}
						&&								&&	\(^{137}\)Ba\(^1\)H					&&	1-0 (22.5)						&&			 59	&&	0.005	\\
\noalign{\vskip 0.5mm}
						&&								&&	\(^{136}\)Ba\(^1\)H					&&	1-0 (22.5)						&&			 59	&&	0.005	\\						
\noalign{\vskip 0.5mm}
						&&								&&	\(^{135}\)Ba\(^1\)H					&&	1-0 (19.5)						&&			 40	&&	0.005	\\						
\hline \hline
\end{tabular}}
\end{table*}
\begin{table*}
\centering
\caption{The MLR$_{p,q}^{r_{ref}}$($N_{\beta}$) potential parameters obtained through a fit to the {\it ab initio} potential, and after initial optimisation with spectral data using \texttt{DPotFit}. All energies in cm$^{-1}$.}

\label{DPotParam}
{
\begin{tabular}{crclrc|ccrclr}
\noalign{\vskip 3mm}
\hline  \hline \noalign{\vskip 1.2mm}
\multicolumn{6}{c|}{MLR$_{8,3}^{3.34}$($12$), {\it ab initio} only} 		   &\multicolumn{6}{c}{MLR$_{8,3}^{3.34}$($12$), {\it ab initio} + spectra}				\\
\noalign{\vskip 0.6mm}
\hline
\noalign{\vskip 1.0mm}
\multicolumn{3}{r}{$\mathfrak{D}_e$ }		& \multicolumn{3}{l|}{16891.5947}	& \multicolumn{3}{r}{$\mathfrak{D}_e$ } 	&  \multicolumn{3}{l|} {16841.8651}		\\
\multicolumn{3}{r}{ r$_e$ }		& \multicolumn{3}{l|}{ 2.2319 {\AA}}	& \multicolumn{3}{r}{ r$_e$ }			& \multicolumn{3}{l}{2.2318 {\AA}}	\\
\multicolumn{3}{r}{ $\overline{dd}$ }		& \multicolumn{3}{l|}{266.447}	& \multicolumn{3}{r}{ $\overline{dd}$ }		& \multicolumn{3}{l}{6.083}		\\
\hline\noalign{\vskip 1.0mm}
$\beta_0$ &    0.114040 	&& $ \beta_7$   &  -87.820760 	&&& $\beta_0$ &   0.096691	&& $\beta_7$	&  -86.606654	\\
$\beta_1$ &   -7.619529		&& $ \beta_8$   & -171.874803	&&& $\beta_1$ &  -7.616896	&& $\beta_8$	& -154.070717	\\
$\beta_2$ &  -17.121381	 	&& $ \beta_9$   &   -5.387603	&&& $\beta_2$ & -16.467554	&& $\beta_9$  	&   34.3428882	\\
$\beta_3$ &  -19.233860 	&& $\beta_{10}$ &  256.820563	&&& $\beta_3$ & -16.902752	&& $\beta_{10}$	&  324.143391	\\
$\beta_4$ &    3.673494		&& $\beta_{11}$ &  260.900999	&&& $\beta_4$ &   6.565879	&& $\beta_{11}$	&  334.502110	\\
$\beta_5$ &   44.305094		&& $\beta_{12}$ &   83.391642	&&& $\beta_5$ &  44.183256	&& $\beta_{12}$	&  116.057107	\\
$\beta_6$ &   31.801460		&&	&							&&& $\beta_6$ &  28.715007	\\

\noalign{\vskip 1mm}
\hline \hline 
\noalign{\vskip 1mm}
\end{tabular}
}

\end{table*}

As \texttt{DPotFit} tries to fit to very accurate spectroscopic data, the initial uncertainty in the fit is very high. Allowing $\beta$ parameters to become fitted parameters produces a final potential which matches the experimental data much more closely than the {\it ab initio} potential alone. During the fitting, both $C_6$ and $C_8$ were fixed at the recommended theoretical values, and were incorporated as the long range tail of the MLR potential with Douketis-type damping \cite{Douketis1982} with \(s=-1\). Additionally, the effect of spin rotation coupling is incorporated through a modification to the MLR potential\cite{LeRoy2013a} using the expansion coefficients listed in Table \ref{DPotBOB}. 


As the considered spectral data covers a range of isotopomers of BaH (electronic transitions of $^{138}$BaH and $^{138}$BaD;
infrared transitions of $^{138}$BaH, $^{137}$BaH, $^{136}$BaH, $^{135}$BaH), all of which have very small values of reduced mass, it is valuable to consider the impact of the Born-Oppenheimer breakdown (BOB)\cite{LeRoy2002} corrections. \texttt{DPotFit} incorporates BOB corrections
through the atom-dependant potential, with the effects parametrised within the radial strength functions: 
\begin{equation}
\tilde{S}_{\text{ad}}^{\text{A}} (r) = y_{p_\text{ad}}^{eq} (r) \,  u_{\infty}^{\text{A}} + 
	\left[ 1 - y_{p_\text{ad}}^{eq}(r) \right] \sum_{i=0}^{N_\text{ad}^\text{A}} u_i^\text{A}\, y_{p_\text{ad}}^{eq} (r)^i
\end{equation}
\begin{equation}
\tilde{R}_{\text{na}}^{\text{A}} (r) = y_{p_\text{na}}^{eq} (r) \, t_{\infty}^{\text{A}} + 
	\left[ 1 - y_{p_\text{na}}^{eq}(r) \right] \sum_{i=0}^{N_\text{na}^\text{A}} t_i^\text{A}\, y_{p_\text{ad}}^{eq} (r)^i
\end{equation}
Where $\tilde{S}_{\text{ad}}^{\text{A}} (r)$ is the atom-dependant adiabatic BOB radial strength function and $\tilde{R}_{\text{na}}^{\text{A}} (r)$
 the non-adiabatic BOB radial strength function. 
{Incorporating fitted terms (Table \ref{DPotBOB}) for $N_\text{ad} = N_\text{na}=2$ along with fitted MLR parameters (Table \ref{DPotParam}) while keeping \(r_e\) fixed at 2.2319 \AA} produced an MLR potential with $\overline{dd} = 6.143$. Allowing \(r_e\) to also be a fitted parameter lowered $\mathfrak{D}_e$ by nearly 50 cm$^{-1}$ but further reduced \(\overline{dd}\) to 6.083. While substantially higher than what would usually be deemed a ``good" fit\cite{LeRoy2013a} ($\overline{dd} <1$), comparison with results from the most recent study\cite{Bernath2013} of the E$(^2\Pi$) $\rightarrow$ X$(^2\Sigma^+$) transition, a data set not used in the \texttt{DPotFit} analysis, highlights the much improved agreement with the experimental observations (Table \ref{DPotFit}) than the raw {\it ab initio} results from Table \ref{bahX-results}.

	%
\begin{figure}
\centering
\includegraphics[scale=0.08]{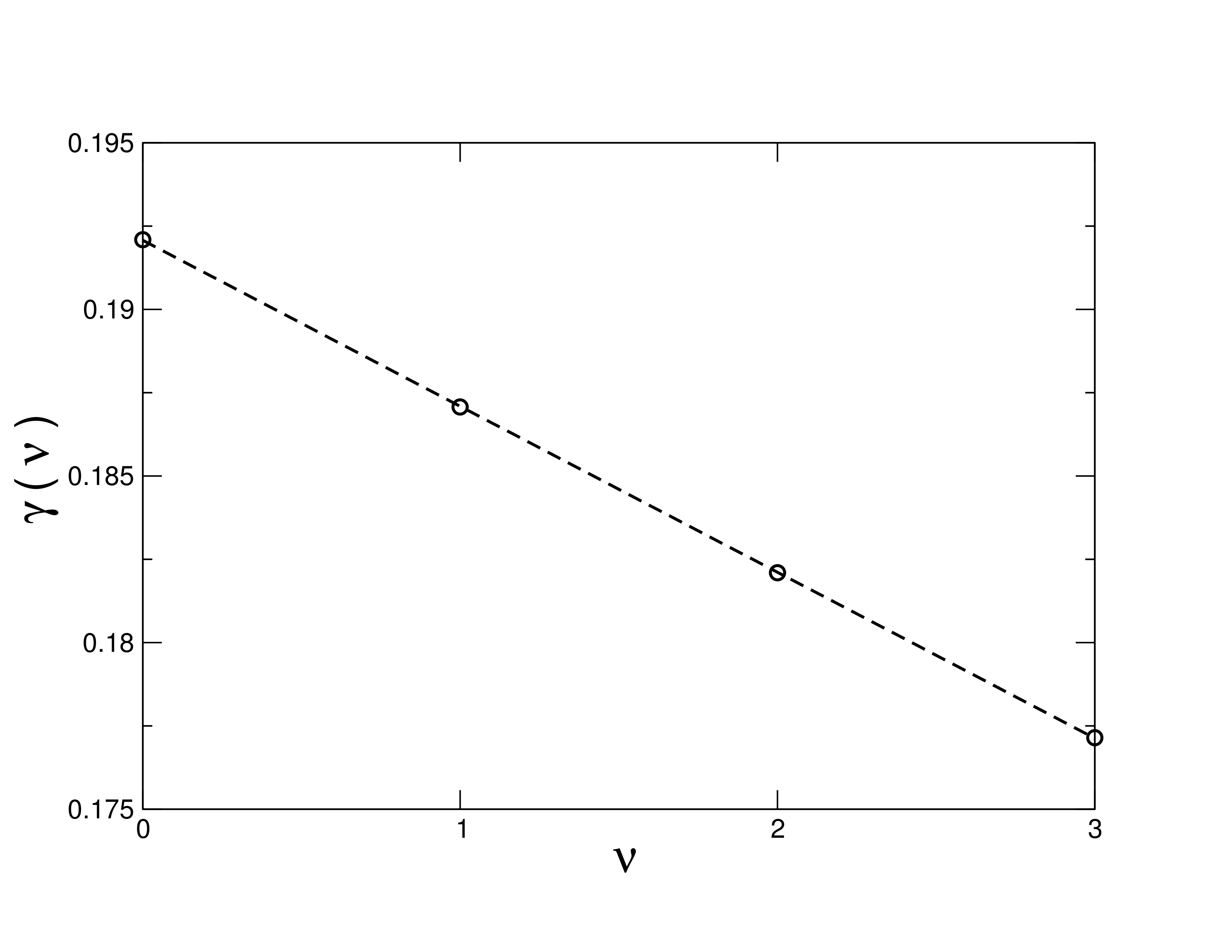}
\caption{The $v$ dependence of the experimental spin-rotation constants $\gamma$ from the experimental work\cite{Bernath1993} of Walker, Hedderich and Bernath.} \label{spinrotation}
\end{figure}
	%
	%
	%

%

\begin{table}
\centering
\caption{Additional parameters defining the $^2\Sigma^+$ spin-rotation coupling and the BOB corrections in the final MLR potential, obtained after initial optimisation with spectral data using \texttt{DPotFit}.}
\vskip 2mm
\label{DPotBOB}

\begin{tabular}{cccc}
\hline  \hline \noalign{\vskip 1.2mm}
\multicolumn{4}{c}
{MLR}$^{3.34}_{8,3}$(12), Additional parameters									\\
\hline  \noalign{\vskip 1.5mm}
			&	 $q^\Sigma$ 	&	 3	 				& 			\\
\noalign{\vskip 0.6mm}
			&	 $w^\Sigma_0$ 	&	 0.058666 				& 			\\
			&	 $w^\Sigma_1$ 	&	-0.034468 				& 			\\
			&	 $w^\Sigma_2$ 	&	-0.015660 				& 			\\
			&				&						&			\\
$p_\text{ad}$	&	6			&	$q_\text{ad}$			&	6		\\
\noalign{\vskip 0.6mm}
$u_0^\text{Ba}$	&	  14.158630		&	$u_0^\text{H}$			&	 -81.921108		\\
$u_1^\text{Ba}$	&	-106.246129		&	$u_1^\text{H}$			&	 244.705054		\\
$u_2^\text{Ba}$	&	  -3.266579		&	$u_2^\text{H}$			&	 -25.518624		\\
$u_{\infty}^\text{Ba}$	&0.000000		&	$u_{\infty}^\text{H}$		&	0.000000	\\
			&				&						&			\\
$p_\text{na}$	&	3			&	$q_\text{na}$			&	3		\\
\noalign{\vskip 0.6mm}
$t_0^\text{Ba}$	&	-0.007379		&	$t_0^\text{H}$			&	 0.007367	\\
$t_1^\text{Ba}$	&	 0.003445		&	$t_1^\text{H}$			&	 0.000210	\\
$t_2^\text{Ba}$	&   -0.015764		&	$t_2^\text{H}$			&	-0.049889	\\
$t_{\infty}^\text{Ba}$	&0.000000		&	$t_{\infty}^\text{H}$		&	0.000000	\\
\noalign{\vskip 0.6mm}
\hline \hline
\end{tabular}

\end{table}
	
In addition, a number of further details were evident while developing the hybrid potential. The experimental values\cite{Bernath1993} for the spin-rotation constant $\gamma (v)$ follow a very clear $v$ dependence\cite{Brown1977,Semczuk2013} (Fig. \ref{spinrotation}) that can be modelled as a simple Taylor (linear) expansion:

\begin{equation}
\gamma (v) = \gamma_{0} + \gamma_{1} \left(v + \frac{1}{2} \right) 
\end{equation}
with $\gamma_0 = 0.19456$ cm$^{-1}$ and $\gamma_1 = -4.9813$ x $10^{-3}$ cm$^{-1}$. Finally, the experimental dissociation energy quoted in the literature has its origin in the sudden onset of rapid predissociation in the C$^2\Sigma^+$ $v^{\prime} =$ 1, $N^{\prime} =$ 10 rovibronic level. This level lies 25942.6 $\pm$ 0.5 cm$^{-1}$ above the X$^2\Sigma^+$ minimum. Assuming that this indicates the dissociation limit of the (diabatic) D$^2\Sigma^+$ potential lies above C$^2\Sigma^+$ $v^{\prime} =$ 1, $N^{\prime} =$ 9 (25874.6 $\pm$ 0.3 cm$^{-1}$) and identifying the asymptote as belonging to Ba($^3$D$_1$) and not Ba($^3$D$_3$), this would fix the dissociation energy D$_e$ of the ground state between 16842.6 cm$^{-1}$ and 16910.6 cm$^{-1}$ which is consistent with the present {\it ab initio} value and other recent theoretical \cite{Skripnikov2013} results,  while the \texttt{DPotFit} result is within 1~cm$^{-1}$ of the lower limit. It is therefore tentatively suggested that the correct dissociation limit for the D$^2\Sigma^+$ state is actually Ba($^3$D$_1$) + H($^2$S$_{\frac{1}{2}}$) and at long range ought to be labelled D${\frac{1}{2}}$(1) in Hund's case (c).
%
	
	%
\begin{figure}
\centering
\includegraphics[scale=0.35]{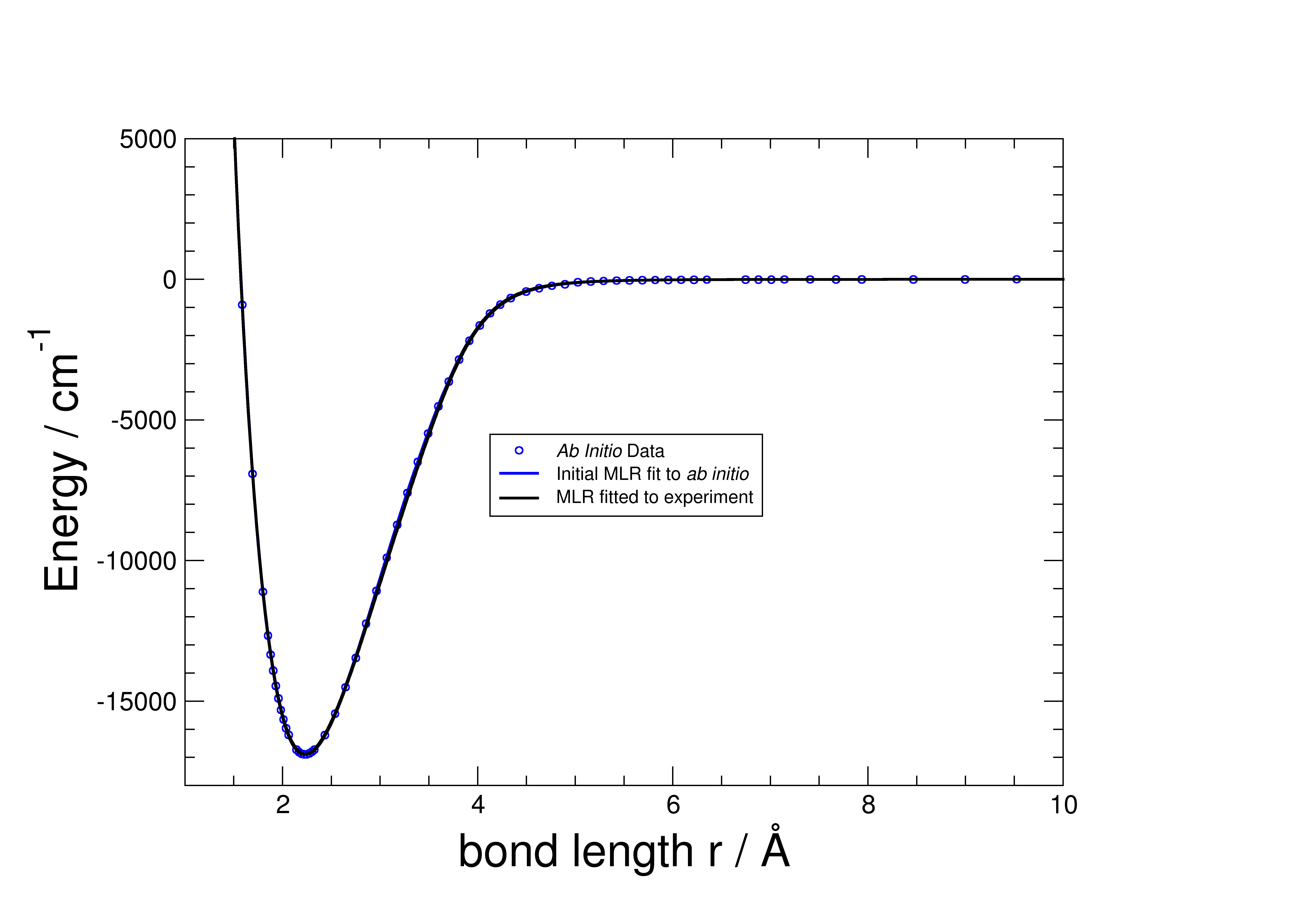}
\caption{MLR$_{8,3}^{3.34}$($12$) fit to {\it ab initio} data points, and final MLR$_{8,3}^{3.34}$($12$) fit to experimental data.} \label{betafit}
\end{figure}

\subsection{Excited states}
\noindent
Extending the {\it ab initio} study to excited states, the lowest energy $^2\Pi$ state was calculated using the even tempered augmented cc-pCV5Z-PP basis set\cite{Peterson2015}. Despite being carried out with a CASSCF calculation including   
5$A_1$, 5$B_1$ and 5$A_2$ states and the MRCI being carried out over 5$B_1$ states, the resulting potential has an r$_e$ of 2.2698~$\text{\AA}$, within 0.38~pm of the best experimental value from Barrow (quoted in Allouche\cite{Allouche1992} {\it et al}). The calculated difference in  X$^2\Sigma^+$ and A$^2\Pi$ bond lengths is therefore $\Delta$r$_e$ = 3.97~pm compared to the experimental difference of 4.16~pm, a discrepancy of just 2 m$\text{\AA}$. Of course, for a truly faithful reproduction of the A$^2\Pi$ state not only must spin-orbit coupling be included in the calculation, but also the angular momentum coupling with other electronic states. However, for FC factors the fact that $\Delta$r$_e$ is so close to experiment is reassuring and small energy shifts in general do not have a large effect.

\begin{figure}
\centering
\includegraphics[scale=0.35]{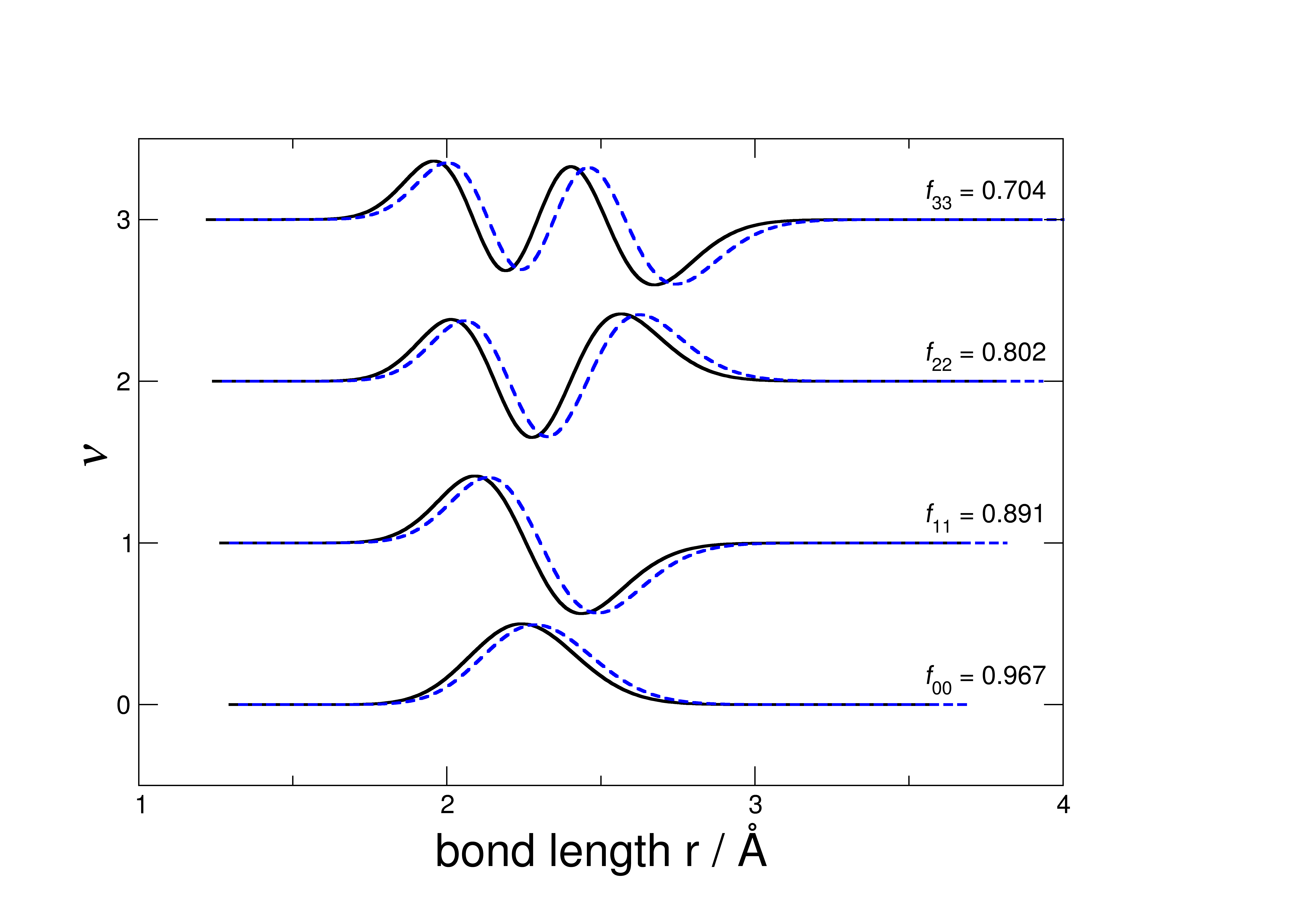}
\caption{The almost identical first four vibrational wavefunctions in the X and A states of BaH. The larger equilibrium bond length displaces the excited state slightly but the overlap is sufficient to ensure diagonal Franck-Condon factors close to 1. Also labelled are the diagonal Franck-Condon (FC) factors $f_{v_{i}v_{i}}$.} \label{diagonal}
\end{figure}

The two potentials were then used to obtain FC factors between the X$^2\Sigma^+$ and A$^2\Pi$ vibronic states using \texttt{LEVEL}.
The highly diagonal nature of the A$^2\Pi$ $ \leftarrow$  X$^2\Sigma^+$ transition is clear when the wavefunctions from both states are compared as in Fig. \ref{diagonal}. The diagonal FC factor $f_{00}$ is almost identical to that found in the earlier study by Lane\cite{Lane2015} using the larger ECP and a much smaller triple zeta quality basis set, though the agreement reduces with each increase in vibrational quantum number (by $f_{33}$ the difference is 5.8$\%$, compared to 0.1$\%$ for $f_{00}$). Using the revised dissociation energy D$_e$ of the ground state ($<$ 16910.6 $\pm$ 0.5 cm$^{-1}$), the $^3$D - $^1$S separation in barium (9372 cm$^{-1}$), the Term energy of the minimum of the A$^2\Pi$ potential (excluding spin-orbit separation) from Barrow\cite{Bernard1989} and co-workers and finally the X$^2\Sigma^+$ zero point energy, the experimental dissociation energy D$_e$ of the A$^2\Pi$ state is {$<$~16001.3~$\pm$~2.0} cm$^{-1}$ which is in good agreement with our calculated value of 16076.75 cm$^{-1}$. This is further support for our proposed revision of the experimental value for the BaH X$^2\Sigma^+$ dissociation energy D$_e$ to $<$~16910.6 $\pm$ 0.5 cm$^{-1}$.

\begin{figure}
\centering
\includegraphics[scale=0.106]{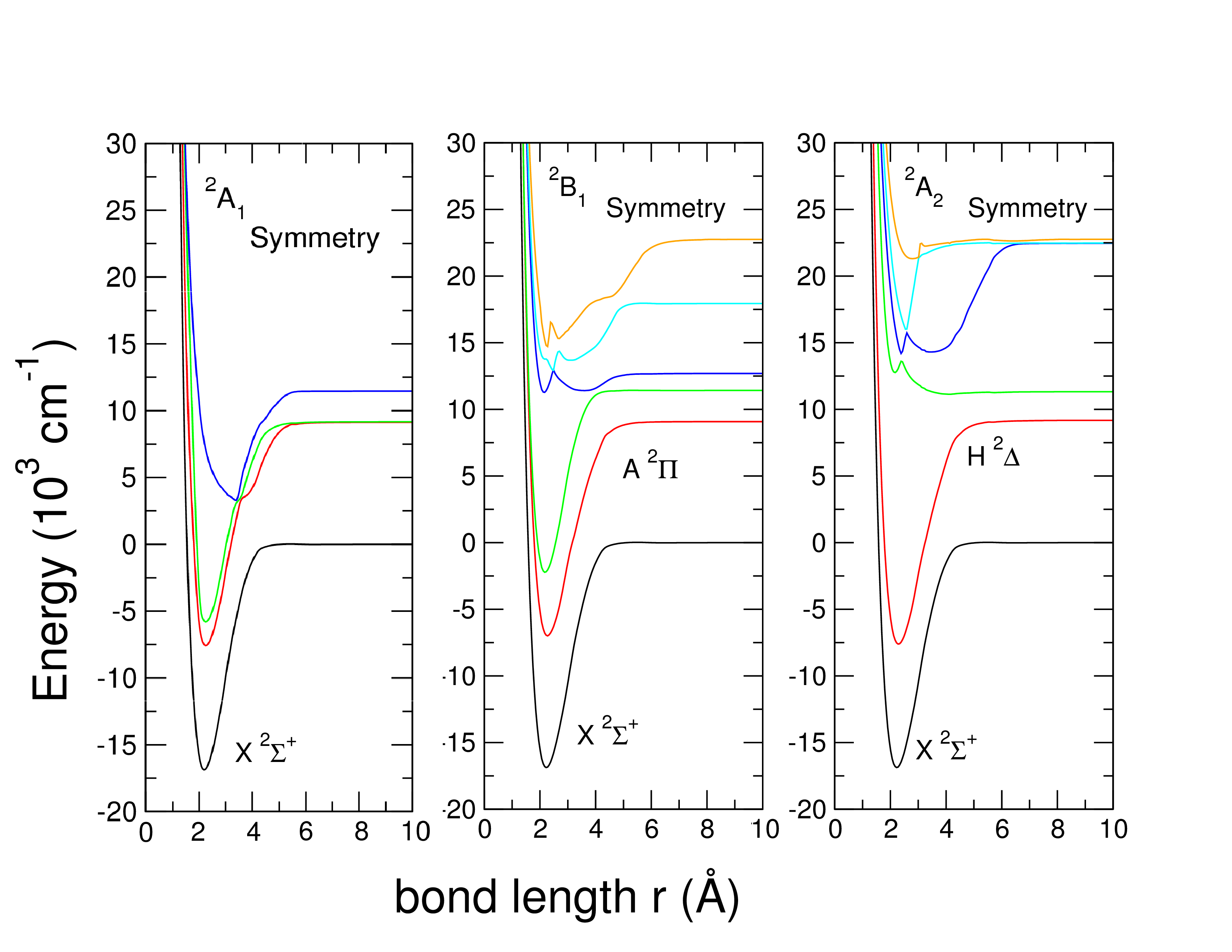}
\caption{Low lying states of BaH using an active 3-electron  model
			   in the MRCI+Q approximation.  The even tempered aug-cc-pCV5Z-PP basis set 
			 is used for the calculations.} \label{15state}
\end{figure}

Using the same basis set, {\it ab initio} calculations 
were performed in the MRCI+Q approximation for five states of each of the $^2A_1$, $^2B_1$ and $^2A_2$ symmetries 
 for bond length out to 40 a$_0$.	
The results from all these calculations are presented in Fig. \ref{15state}. 
The $^2A_1$ states are the most common symmetry at lower energies and therefore valence states dominate in Fig. \ref{15state}. Comparing to the previous AV6Z calculation, the less diffuse nature of the largest Gaussian functions in our adopted aug-pCV$n$Z-PP basis sets has resulted in the greater weighting of valence states over Rydbergs and so the C$^2\Sigma^+$ state has disappeared. To describe the Rydberg character of the 
$^2\Sigma^+$ states, it's necessary to expand the active space in the MRCI+Q 
calculations to at least (8$a_1$, 5$b_1$, 3$b_2$, 2$_2$).  
\begin{table}
\centering
\caption{Vibrational term energies in the X$^2\Sigma^+$ ground state of the BaH molecule using \texttt{DPotFit}. Comparison of the vibrational levels from the present combined {\it ab initio}/spectroscopic potential and experimental\cite{Bernath2013} data {independent of fit}. All energies in cm$^{-1}$.}
              
\label{DPotFit}

\begin{tabular}{rcD{.}{.}{3}ccD{.}{.}{3}ccD{.}{.}{3}ccD{.}{.}{3}ccD{.}{.}{3}}
\noalign{\vskip 2mm}
\hline \hline \noalign{\vskip 1.3mm} 
$v$ && \multicolumn{1}{c}{Experiment} &&& \multicolumn{1}{c}{\texttt{DPotFit}\(^a\)}  &&& \multicolumn{1}{c}{$\Delta$E} &&& \multicolumn{1}{c}{\texttt{DPotFit}\(^b\)}	&&& \multicolumn{1}{c}{$\Delta$E}  \\
    && \multicolumn{1}{c}{(Ram {\it et al})}	&&& \multicolumn{1}{c}{\(r_e\) fix}	  &&& \multicolumn{1}{c}{\(r_e\) fix} &&& 	\multicolumn{1}{c}{\(r_e\) fit}		 &&& \multicolumn{1}{c}{\(r_e\) fit}   \\
\hline
\noalign{\vskip 2mm}
0 &&	   0.0000	&&&	   0.0000	&&&	0.0000		&&&	 0.0000 &&&	 0.0000\\
1 &&	1139.2896 	&&&	1139.2884	&&& -0.0012	&&&	1139.2897	&&&	 0.0001\\
2 &&	2249.6062	&&&	2249.6084	&&&	-0.0022	&&&	2249.6078	&&& -0.0016\\
3 &&	3331.1192	&&&	3331.1212	&&&	-0.0020	&&&	3331.1234	&&&	-0.0042\\
\noalign{\vskip 1mm}
\hline \hline
\end{tabular}
\begin{flushleft}
\(^a\) Zero Point Energy = 580.6050~cm\(^{-1}\)\\
\(^b\) Zero Point Energy = 580.5910~cm\(^{-1}\)
\end{flushleft}

\begin{flushleft}
\end{flushleft}
\end{table}

  	%
	%
	%

	%
	%

\section{Summary and Conclusion}
\noindent
In this paper we have attempted to develop a reliable potential energy curve for the X$^2\Sigma^+$ state of the BaH radical. Encouraged by the close agreement in the literature between relativistic and non-relativistic {\it ab initio} calculations, we have used the MRCI+Q method to determine the shape of the X$^2\Sigma^+$ potential and the dissociation energy. With an even-tempered augmented cc-pCV$n$Z-PP basis set taken to the CBS limit, we have calculated the equilibrium bond length to within 0.031~pm of the experimental value and of comparable accuracy to the more trusted CCSD(T) technique. While the ten 5$s$5$p$6$s$ electrons from barium are necessary to calculate the static correlation energy, only the valence 6$s$ electrons are needed to cover the majority of the dynamic correlation. The computed dissociation energy (16895.12~cm$^{-1}$) is in good agreement with other theoretical values suggesting a revision of the experimental dissociation energy to D$_e$ $<$16910.6~cm$^{-1}$. The calculated dissociation energy of the excited A$^2\Pi$ state (16076.75~cm$^{-1}$) is also consistent with this experimental value. The vibrational levels of the {\it ab initio} X$^2\Sigma^+$  potential achieve an average agreement of just over 1 cm$^{-1}$ with the best experimental values. Refining the potential with experimental data using \texttt{DPotFit} improves the agreement to better than 0.005~cm$^{-1}$ 
 for the lowest vibrational levels.       

\section*{acknowledgments}
\noindent
I. C. L. thanks the Leverhulme Trust for financial support, a PDRA fellowship 
for B. M. M. and a Ph D studentship for K. M. under Research Grant RPG-2014-212. 
We would like to thank Professor K. Peterson for providing the explicitly optimized 
augmented functions for the barium atom for his cc-pCV$n$Z-PP basis sets and 
for many helpful discussions regarding this {\it ab initio} work. 
Grants of computational time at the National Energy Research 
Scientific Computing (NERSC) Center in Oakland, CA, USA and at the
 High Performance Computing Center Stuttgart (HLRS) of the University of Stuttgart, 
Stuttgart, Germany, where some of these computations were performed, are gratefully acknowledged.
%
%
%
%
%
\bibliographystyle{apsrev4-1}
\bibliography{refs}

\begin{thebibliography}{72}%
\makeatletter
\providecommand \@ifxundefined [1]{%
 \@ifx{#1\undefined}
}%
\providecommand \@ifnum [1]{%
 \ifnum #1\expandafter \@firstoftwo
 \else \expandafter \@secondoftwo
 \fi
}%
\providecommand \@ifx [1]{%
 \ifx #1\expandafter \@firstoftwo
 \else \expandafter \@secondoftwo
 \fi
}%
\providecommand \natexlab [1]{#1}%
\providecommand \enquote  [1]{``#1''}%
\providecommand \bibnamefont  [1]{#1}%
\providecommand \bibfnamefont [1]{#1}%
\providecommand \citenamefont [1]{#1}%
\providecommand \href@noop [0]{\@secondoftwo}%
\providecommand \href [0]{\begingroup \@sanitize@url \@href}%
\providecommand \@href[1]{\@@startlink{#1}\@@href}%
\providecommand \@@href[1]{\endgroup#1\@@endlink}%
\providecommand \@sanitize@url [0]{\catcode `\\12\catcode `\$12\catcode
  `\&12\catcode `\#12\catcode `\^12\catcode `\_12\catcode `\%12\relax}%
\providecommand \@@startlink[1]{}%
\providecommand \@@endlink[0]{}%
\providecommand \url  [0]{\begingroup\@sanitize@url \@url }%
\providecommand \@url [1]{\endgroup\@href {#1}{\urlprefix }}%
\providecommand \urlprefix  [0]{URL }%
\providecommand \Eprint [0]{\href }%
\providecommand \doibase [0]{http://dx.doi.org/}%
\providecommand \selectlanguage [0]{\@gobble}%
\providecommand \bibinfo  [0]{\@secondoftwo}%
\providecommand \bibfield  [0]{\@secondoftwo}%
\providecommand \translation [1]{[#1]}%
\providecommand \BibitemOpen [0]{}%
\providecommand \bibitemStop [0]{}%
\providecommand \bibitemNoStop [0]{.\EOS\space}%
\providecommand \EOS [0]{\spacefactor3000\relax}%
\providecommand \BibitemShut  [1]{\csname bibitem#1\endcsname}%
\let\auto@bib@innerbib\@empty
\bibitem [{\citenamefont {Lane}(2015)}]{Lane2015}%
  \BibitemOpen
  \bibfield  {author} {\bibinfo {author} {\bibfnamefont {I.~C.}\ \bibnamefont
  {Lane}},\ }\href@noop {} {\bibfield  {journal} {\bibinfo  {journal} {Phys.
  Rev. A}\ }\textbf {\bibinfo {volume} {92}},\ \bibinfo {pages} {022511}
  (\bibinfo {year} {2015})}\BibitemShut {NoStop}%
\bibitem [{\citenamefont {Werner}\ and\ \citenamefont
  {Knowles}(1985{\natexlab{a}})}]{Knowles1985}%
  \BibitemOpen
  \bibfield  {author} {\bibinfo {author} {\bibfnamefont {H.~J.}\ \bibnamefont
  {Werner}}\ and\ \bibinfo {author} {\bibfnamefont {P.~J.}\ \bibnamefont
  {Knowles}},\ }\href@noop {} {\bibfield  {journal} {\bibinfo  {journal} {Chem.
  Phys. Lett.}\ }\textbf {\bibinfo {volume} {\textbf{115}}},\ \bibinfo {pages}
  {259} (\bibinfo {year} {1985}{\natexlab{a}})}\BibitemShut {NoStop}%
\bibitem [{\citenamefont {Werner}\ and\ \citenamefont
  {Knowles}(1988)}]{Knowles1988}%
  \BibitemOpen
  \bibfield  {author} {\bibinfo {author} {\bibfnamefont {H.~J.}\ \bibnamefont
  {Werner}}\ and\ \bibinfo {author} {\bibfnamefont {P.~J.}\ \bibnamefont
  {Knowles}},\ }\href@noop {} {\bibfield  {journal} {\bibinfo  {journal} {J.
  Chem. Phys.}\ }\textbf {\bibinfo {volume} {\textbf{89}}},\ \bibinfo {pages}
  {5803} (\bibinfo {year} {1988})}\BibitemShut {NoStop}%
\bibitem [{\citenamefont {Tarallo}\ \emph {et~al.}(2016)\citenamefont
  {Tarallo}, \citenamefont {Iwata},\ and\ \citenamefont
  {Zelevinsky}}]{Tarallo2016}%
  \BibitemOpen
  \bibfield  {author} {\bibinfo {author} {\bibfnamefont {M.~G.}\ \bibnamefont
  {Tarallo}}, \bibinfo {author} {\bibfnamefont {G.~Z.}\ \bibnamefont {Iwata}},
  \ and\ \bibinfo {author} {\bibfnamefont {T.}~\bibnamefont {Zelevinsky}},\
  }\href@noop {} {\bibfield  {journal} {\bibinfo  {journal} {Phys. Rev. A}\
  }\textbf {\bibinfo {volume} {\textbf{93}}},\ \bibinfo {pages} {032509}
  (\bibinfo {year} {2016})}\BibitemShut {NoStop}%
\bibitem [{\citenamefont {Ram}\ and\ \citenamefont
  {Bernath}(2013)}]{Bernath2013}%
  \BibitemOpen
  \bibfield  {author} {\bibinfo {author} {\bibfnamefont {R.~S.}\ \bibnamefont
  {Ram}}\ and\ \bibinfo {author} {\bibfnamefont {P.~F.}\ \bibnamefont
  {Bernath}},\ }\href@noop {} {\bibfield  {journal} {\bibinfo  {journal} {J.
  Mol. Spec.}\ }\textbf {\bibinfo {volume} {\textbf{283}}},\ \bibinfo {pages}
  {18} (\bibinfo {year} {2013})}\BibitemShut {NoStop}%
\bibitem [{\citenamefont {Fuentealba}\ \emph {et~al.}(1987)\citenamefont
  {Fuentealba}, \citenamefont {Reyes}, \citenamefont {Stoll},\ and\
  \citenamefont {Preuss}}]{Preuss1987}%
  \BibitemOpen
  \bibfield  {author} {\bibinfo {author} {\bibfnamefont {P.}~\bibnamefont
  {Fuentealba}}, \bibinfo {author} {\bibfnamefont {O.}~\bibnamefont {Reyes}},
  \bibinfo {author} {\bibfnamefont {H.}~\bibnamefont {Stoll}}, \ and\ \bibinfo
  {author} {\bibfnamefont {H.}~\bibnamefont {Preuss}},\ }\href@noop {}
  {\bibfield  {journal} {\bibinfo  {journal} {J. Chem. Phys.}\ }\textbf
  {\bibinfo {volume} {\textbf{87}}},\ \bibinfo {pages} {5338} (\bibinfo {year}
  {1987})}\BibitemShut {NoStop}%
\bibitem [{\citenamefont {Kaupp}\ \emph {et~al.}(1991)\citenamefont {Kaupp},
  \citenamefont {Schleyer}, \citenamefont {Stoll},\ and\ \citenamefont
  {Preuss}}]{Kaupp1991}%
  \BibitemOpen
  \bibfield  {author} {\bibinfo {author} {\bibfnamefont {M.}~\bibnamefont
  {Kaupp}}, \bibinfo {author} {\bibfnamefont {P.}~\bibnamefont {Schleyer}},
  \bibinfo {author} {\bibfnamefont {H.}~\bibnamefont {Stoll}}, \ and\ \bibinfo
  {author} {\bibfnamefont {H.}~\bibnamefont {Preuss}},\ }\href@noop {}
  {\bibfield  {journal} {\bibinfo  {journal} {J. Chem. Phys.}\ }\textbf
  {\bibinfo {volume} {\textbf{94}}},\ \bibinfo {pages} {1360} (\bibinfo {year}
  {1991})}\BibitemShut {NoStop}%
\bibitem [{\citenamefont {Allouche}\ \emph {et~al.}(1992)\citenamefont
  {Allouche}, \citenamefont {Nicolas}, \citenamefont {Barthelat},\ and\
  \citenamefont {Spiegelmann}}]{Allouche1992}%
  \BibitemOpen
  \bibfield  {author} {\bibinfo {author} {\bibfnamefont {A.~R.}\ \bibnamefont
  {Allouche}}, \bibinfo {author} {\bibfnamefont {G.}~\bibnamefont {Nicolas}},
  \bibinfo {author} {\bibfnamefont {J.~C.}\ \bibnamefont {Barthelat}}, \ and\
  \bibinfo {author} {\bibfnamefont {F.}~\bibnamefont {Spiegelmann}},\
  }\href@noop {} {\bibfield  {journal} {\bibinfo  {journal} {J. Chem. Phys.}\
  }\textbf {\bibinfo {volume} {\textbf{96}}},\ \bibinfo {pages} {7646}
  (\bibinfo {year} {1992})}\BibitemShut {NoStop}%
\bibitem [{\citenamefont {Skripnikov}\ \emph {et~al.}(2013)\citenamefont
  {Skripnikov}, \citenamefont {Mosyagin},\ and\ \citenamefont
  {Titov}}]{Skripnikov2013}%
  \BibitemOpen
  \bibfield  {author} {\bibinfo {author} {\bibfnamefont {L.~V.}\ \bibnamefont
  {Skripnikov}}, \bibinfo {author} {\bibfnamefont {N.~S.}\ \bibnamefont
  {Mosyagin}}, \ and\ \bibinfo {author} {\bibfnamefont {A.~V.}\ \bibnamefont
  {Titov}},\ }\href@noop {} {\bibfield  {journal} {\bibinfo  {journal} {Chem.
  Phys. Lett.}\ }\textbf {\bibinfo {volume} {\textbf{555}}},\ \bibinfo {pages}
  {79} (\bibinfo {year} {2013})}\BibitemShut {NoStop}%
\bibitem [{\citenamefont {Kostelecky}\ and\ \citenamefont
  {Vargas}(2015)}]{Kostelecky2015}%
  \BibitemOpen
  \bibfield  {author} {\bibinfo {author} {\bibfnamefont {V.~A.}\ \bibnamefont
  {Kostelecky}}\ and\ \bibinfo {author} {\bibfnamefont {A.~J.}\ \bibnamefont
  {Vargas}},\ }\href@noop {} {\bibfield  {journal} {\bibinfo  {journal} {Phys.
  Rev. D}\ }\textbf {\bibinfo {volume} {92}},\ \bibinfo {pages} {056002}
  (\bibinfo {year} {2015})}\BibitemShut {NoStop}%
\bibitem [{\citenamefont {Ubachs}\ \emph {et~al.}(2016)\citenamefont {Ubachs},
  \citenamefont {Koelemeij}, \citenamefont {Eikema},\ and\ \citenamefont
  {Salumbides}}]{Ubachs2016}%
  \BibitemOpen
  \bibfield  {author} {\bibinfo {author} {\bibfnamefont {W.}~\bibnamefont
  {Ubachs}}, \bibinfo {author} {\bibfnamefont {J.}~\bibnamefont {Koelemeij}},
  \bibinfo {author} {\bibfnamefont {K.}~\bibnamefont {Eikema}}, \ and\ \bibinfo
  {author} {\bibfnamefont {E.}~\bibnamefont {Salumbides}},\ }\href@noop {}
  {\bibfield  {journal} {\bibinfo  {journal} {J. Mol. Spectro.}\ }\textbf
  {\bibinfo {volume} {320}},\ \bibinfo {pages} {1} (\bibinfo {year}
  {2016})}\BibitemShut {NoStop}%
\bibitem [{\citenamefont {Rigden}(1983)}]{Rigden1983}%
  \BibitemOpen
  \bibfield  {author} {\bibinfo {author} {\bibfnamefont {J.~S.}\ \bibnamefont
  {Rigden}},\ }\href@noop {} {\bibfield  {journal} {\bibinfo  {journal} {Hist.
  Stud. Phys. Sci.}\ }\textbf {\bibinfo {volume} {13}},\ \bibinfo {pages} {335}
  (\bibinfo {year} {1983})}\BibitemShut {NoStop}%
\bibitem [{\citenamefont {Kragh}(1985)}]{Kragh1985}%
  \BibitemOpen
  \bibfield  {author} {\bibinfo {author} {\bibfnamefont {H.}~\bibnamefont
  {Kragh}},\ }\href@noop {} {\bibfield  {journal} {\bibinfo  {journal} {Hist.
  Stud. Phys. Sci.}\ }\textbf {\bibinfo {volume} {15}},\ \bibinfo {pages} {67}
  (\bibinfo {year} {1985})}\BibitemShut {NoStop}%
\bibitem [{\citenamefont {Cagnac}\ \emph {et~al.}(1994)\citenamefont {Cagnac},
  \citenamefont {Plimmer}, \citenamefont {Julien},\ and\ \citenamefont
  {Biraben}}]{Cagnac1994}%
  \BibitemOpen
  \bibfield  {author} {\bibinfo {author} {\bibfnamefont {B.}~\bibnamefont
  {Cagnac}}, \bibinfo {author} {\bibfnamefont {M.~D.}\ \bibnamefont {Plimmer}},
  \bibinfo {author} {\bibfnamefont {L.}~\bibnamefont {Julien}}, \ and\ \bibinfo
  {author} {\bibfnamefont {F.}~\bibnamefont {Biraben}},\ }\href@noop {}
  {\bibfield  {journal} {\bibinfo  {journal} {Rep. Prog. Phys.}\ }\textbf
  {\bibinfo {volume} {57}},\ \bibinfo {pages} {853} (\bibinfo {year}
  {1994})}\BibitemShut {NoStop}%
\bibitem [{\citenamefont {Karshenboim}(2015)}]{Karshenboim2015}%
  \BibitemOpen
  \bibfield  {author} {\bibinfo {author} {\bibfnamefont {S.~G.}\ \bibnamefont
  {Karshenboim}},\ }\href@noop {} {\bibfield  {journal} {\bibinfo  {journal}
  {Phys. Rev. A}\ }\textbf {\bibinfo {volume} {91}},\ \bibinfo {pages} {012515}
  (\bibinfo {year} {2015})}\BibitemShut {NoStop}%
\bibitem [{\citenamefont {Koput}(2011)}]{Koput2011}%
  \BibitemOpen
  \bibfield  {author} {\bibinfo {author} {\bibfnamefont {J.}~\bibnamefont
  {Koput}},\ }\href@noop {} {\bibfield  {journal} {\bibinfo  {journal} {J.
  Chem. Phys.}\ }\textbf {\bibinfo {volume} {\textbf{135}}},\ \bibinfo {pages}
  {244308} (\bibinfo {year} {2011})}\BibitemShut {NoStop}%
\bibitem [{\citenamefont {Dattani}(2015)}]{Dattani2015}%
  \BibitemOpen
  \bibfield  {author} {\bibinfo {author} {\bibfnamefont {N.~S.}\ \bibnamefont
  {Dattani}},\ }\href@noop {} {\bibfield  {journal} {\bibinfo  {journal} {J.
  Mol. Spectrosc.}\ }\textbf {\bibinfo {volume} {\textbf{311}}},\ \bibinfo
  {pages} {76} (\bibinfo {year} {2015})}\BibitemShut {NoStop}%
\bibitem [{\citenamefont {Werner}\ \emph {et~al.}()\citenamefont {Werner},
  \citenamefont {Knowles}, \citenamefont {Lindh}, \citenamefont {Manby},\ and\
  \citenamefont {{\it et. al.}}}]{Werner2010}%
  \BibitemOpen
  \bibfield  {author} {\bibinfo {author} {\bibfnamefont {H.-J.}\ \bibnamefont
  {Werner}}, \bibinfo {author} {\bibfnamefont {P.~J.}\ \bibnamefont {Knowles}},
  \bibinfo {author} {\bibfnamefont {R.}~\bibnamefont {Lindh}}, \bibinfo
  {author} {\bibfnamefont {F.~R.}\ \bibnamefont {Manby}}, \ and\ \bibinfo
  {author} {\bibfnamefont {M.~S.}\ \bibnamefont {{\it et. al.}}},\ }\href
  {http://www.molpro.net} {\enquote {\bibinfo {title} {{MOLPRO 2010,
  http://www.molpro.net}},}\ }\BibitemShut {NoStop}%
\bibitem [{\citenamefont {Gdanitz}\ and\ \citenamefont
  {Ahlrichs}(1988)}]{Gdanitz1988}%
  \BibitemOpen
  \bibfield  {author} {\bibinfo {author} {\bibfnamefont {R.~J.}\ \bibnamefont
  {Gdanitz}}\ and\ \bibinfo {author} {\bibfnamefont {R.}~\bibnamefont
  {Ahlrichs}},\ }\href@noop {} {\bibfield  {journal} {\bibinfo  {journal}
  {Chem. Phys. Lett.}\ }\textbf {\bibinfo {volume} {\textbf{143}}},\ \bibinfo
  {pages} {413} (\bibinfo {year} {1988})}\BibitemShut {NoStop}%
\bibitem [{\citenamefont {{Le Roy}}\ \emph
  {et~al.}(2006{\natexlab{a}})\citenamefont {{Le Roy}}, \citenamefont
  {Appadoo}, \citenamefont {Colin},\ and\ \citenamefont {Bernath}}]{Leroy2006}%
  \BibitemOpen
  \bibfield  {author} {\bibinfo {author} {\bibfnamefont {R.~J.}\ \bibnamefont
  {{Le Roy}}}, \bibinfo {author} {\bibfnamefont {D.~R.}\ \bibnamefont
  {Appadoo}}, \bibinfo {author} {\bibfnamefont {R.}~\bibnamefont {Colin}}, \
  and\ \bibinfo {author} {\bibfnamefont {P.~F.}\ \bibnamefont {Bernath}},\
  }\href@noop {} {\bibfield  {journal} {\bibinfo  {journal} {J. Mol.
  Spectrosc.}\ }\textbf {\bibinfo {volume} {\textbf{236}}},\ \bibinfo {pages}
  {178} (\bibinfo {year} {2006}{\natexlab{a}})}\BibitemShut {NoStop}%
\bibitem [{\citenamefont {Kopp}\ \emph
  {et~al.}(1966{\natexlab{a}})\citenamefont {Kopp}, \citenamefont {$\rm
  \AA$slund}, \citenamefont {Edvinsson},\ and\ \citenamefont
  {Lindgren}}]{Lindgren1965}%
  \BibitemOpen
  \bibfield  {author} {\bibinfo {author} {\bibfnamefont {I.}~\bibnamefont
  {Kopp}}, \bibinfo {author} {\bibfnamefont {N.}~\bibnamefont {$\rm
  \AA$slund}}, \bibinfo {author} {\bibfnamefont {G.}~\bibnamefont {Edvinsson}},
  \ and\ \bibinfo {author} {\bibfnamefont {B.}~\bibnamefont {Lindgren}},\
  }\href@noop {} {\bibfield  {journal} {\bibinfo  {journal} {Ark. Fysik.}\
  }\textbf {\bibinfo {volume} {\textbf{30}}},\ \bibinfo {pages} {321} (\bibinfo
  {year} {1966}{\natexlab{a}})}\BibitemShut {NoStop}%
\bibitem [{\citenamefont {Eagle}(1909)}]{Eagle1909}%
  \BibitemOpen
  \bibfield  {author} {\bibinfo {author} {\bibfnamefont {A.}~\bibnamefont
  {Eagle}},\ }\href@noop {} {\bibfield  {journal} {\bibinfo  {journal}
  {Astrophys. J.}\ }\textbf {\bibinfo {volume} {\textbf{30}}},\ \bibinfo
  {pages} {231} (\bibinfo {year} {1909})}\BibitemShut {NoStop}%
\bibitem [{\citenamefont {Schaafsma}(1932)}]{Schaafsma1932}%
  \BibitemOpen
  \bibfield  {author} {\bibinfo {author} {\bibfnamefont {A.}~\bibnamefont
  {Schaafsma}},\ }\href@noop {} {\bibfield  {journal} {\bibinfo  {journal} {Z.
  Phys.}\ }\textbf {\bibinfo {volume} {\textbf{74}}},\ \bibinfo {pages} {254}
  (\bibinfo {year} {1932})}\BibitemShut {NoStop}%
\bibitem [{\citenamefont {Fredrickson}\ and\ \citenamefont
  {Watson}(1932)}]{Watson1932}%
  \BibitemOpen
  \bibfield  {author} {\bibinfo {author} {\bibfnamefont {W.~R.}\ \bibnamefont
  {Fredrickson}}\ and\ \bibinfo {author} {\bibfnamefont {W.~W.}\ \bibnamefont
  {Watson}},\ }\href@noop {} {\bibfield  {journal} {\bibinfo  {journal} {Phys.
  Rev.}\ }\textbf {\bibinfo {volume} {\textbf{39}}},\ \bibinfo {pages} {753}
  (\bibinfo {year} {1932})}\BibitemShut {NoStop}%
\bibitem [{\citenamefont {Funke}(1933)}]{Funke1933}%
  \BibitemOpen
  \bibfield  {author} {\bibinfo {author} {\bibfnamefont {G.}~\bibnamefont
  {Funke}},\ }\href@noop {} {\bibfield  {journal} {\bibinfo  {journal} {Z.
  Phys.}\ }\textbf {\bibinfo {volume} {\textbf{84}}},\ \bibinfo {pages} {610}
  (\bibinfo {year} {1933})}\BibitemShut {NoStop}%
\bibitem [{\citenamefont {Watson}(1933)}]{Watson1933}%
  \BibitemOpen
  \bibfield  {author} {\bibinfo {author} {\bibfnamefont {W.~W.}\ \bibnamefont
  {Watson}},\ }\href@noop {} {\bibfield  {journal} {\bibinfo  {journal} {Phys.
  Rev.}\ }\textbf {\bibinfo {volume} {\textbf{43}}},\ \bibinfo {pages} {9}
  (\bibinfo {year} {1933})}\BibitemShut {NoStop}%
\bibitem [{\citenamefont {Watson}(1935)}]{Watson1935}%
  \BibitemOpen
  \bibfield  {author} {\bibinfo {author} {\bibfnamefont {W.~W.}\ \bibnamefont
  {Watson}},\ }\href@noop {} {\bibfield  {journal} {\bibinfo  {journal} {Phys.
  Rev.}\ }\textbf {\bibinfo {volume} {\textbf{47}}},\ \bibinfo {pages} {213}
  (\bibinfo {year} {1935})}\BibitemShut {NoStop}%
\bibitem [{\citenamefont {Koontz}\ and\ \citenamefont
  {Watson}(1935)}]{Koontz1935}%
  \BibitemOpen
  \bibfield  {author} {\bibinfo {author} {\bibfnamefont {P.~G.}\ \bibnamefont
  {Koontz}}\ and\ \bibinfo {author} {\bibfnamefont {W.~W.}\ \bibnamefont
  {Watson}},\ }\href@noop {} {\bibfield  {journal} {\bibinfo  {journal} {Phys.
  Rev.}\ }\textbf {\bibinfo {volume} {\textbf{48}}},\ \bibinfo {pages} {937}
  (\bibinfo {year} {1935})}\BibitemShut {NoStop}%
\bibitem [{\citenamefont {Fabre}\ \emph {et~al.}(1987)\citenamefont {Fabre},
  \citenamefont {El-Hachimi}, \citenamefont {Stringat}, \citenamefont
  {Effantin}, \citenamefont {Bernard}, \citenamefont {d$^{’}$Incan},\ and\
  \citenamefont {Verg{\`e}s}}]{Fabre1987}%
  \BibitemOpen
  \bibfield  {author} {\bibinfo {author} {\bibfnamefont {G.}~\bibnamefont
  {Fabre}}, \bibinfo {author} {\bibfnamefont {A.}~\bibnamefont {El-Hachimi}},
  \bibinfo {author} {\bibfnamefont {R.}~\bibnamefont {Stringat}}, \bibinfo
  {author} {\bibfnamefont {C.}~\bibnamefont {Effantin}}, \bibinfo {author}
  {\bibfnamefont {A.}~\bibnamefont {Bernard}}, \bibinfo {author} {\bibfnamefont
  {J.}~\bibnamefont {d$^{’}$Incan}}, \ and\ \bibinfo {author} {\bibfnamefont
  {J.}~\bibnamefont {Verg{\`e}s}},\ }\href@noop {} {\bibfield  {journal}
  {\bibinfo  {journal} {J. Phys. B: At. Mol. Phys.}\ }\textbf {\bibinfo
  {volume} {\textbf{20}}},\ \bibinfo {pages} {1933} (\bibinfo {year}
  {1987})}\BibitemShut {NoStop}%
\bibitem [{\citenamefont {Grunstr{\"o}m}(1936)}]{Grunstrom1936}%
  \BibitemOpen
  \bibfield  {author} {\bibinfo {author} {\bibfnamefont {B.}~\bibnamefont
  {Grunstr{\"o}m}},\ }\href@noop {} {\bibfield  {journal} {\bibinfo  {journal}
  {Z. Phys.}\ }\textbf {\bibinfo {volume} {\textbf{99}}},\ \bibinfo {pages}
  {595} (\bibinfo {year} {1936})}\BibitemShut {NoStop}%
\bibitem [{\citenamefont {Funke}\ and\ \citenamefont
  {Grunstr{\"o}m}(1936)}]{Funke1936}%
  \BibitemOpen
  \bibfield  {author} {\bibinfo {author} {\bibfnamefont {G.~W.}\ \bibnamefont
  {Funke}}\ and\ \bibinfo {author} {\bibfnamefont {B.}~\bibnamefont
  {Grunstr{\"o}m}},\ }\href@noop {} {\bibfield  {journal} {\bibinfo  {journal}
  {Z. Phys.}\ }\textbf {\bibinfo {volume} {\textbf{100}}},\ \bibinfo {pages}
  {293} (\bibinfo {year} {1936})}\BibitemShut {NoStop}%
\bibitem [{\citenamefont {Edvinsson}\ \emph {et~al.}(1963)\citenamefont
  {Edvinsson}, \citenamefont {Kopp}, \citenamefont {Lindgren},\ and\
  \citenamefont {$\rm \AA$slund}}]{Aslund1963}%
  \BibitemOpen
  \bibfield  {author} {\bibinfo {author} {\bibfnamefont {G.}~\bibnamefont
  {Edvinsson}}, \bibinfo {author} {\bibfnamefont {I.}~\bibnamefont {Kopp}},
  \bibinfo {author} {\bibfnamefont {B.}~\bibnamefont {Lindgren}}, \ and\
  \bibinfo {author} {\bibfnamefont {N.}~\bibnamefont {$\rm \AA$slund}},\
  }\href@noop {} {\bibfield  {journal} {\bibinfo  {journal} {Ark. Fysik.}\
  }\textbf {\bibinfo {volume} {\textbf{25}}},\ \bibinfo {pages} {95} (\bibinfo
  {year} {1963})}\BibitemShut {NoStop}%
\bibitem [{\citenamefont {Kopp}\ and\ \citenamefont
  {Wirhed}(1966)}]{Wirhed1966}%
  \BibitemOpen
  \bibfield  {author} {\bibinfo {author} {\bibfnamefont {I.}~\bibnamefont
  {Kopp}}\ and\ \bibinfo {author} {\bibfnamefont {R.}~\bibnamefont {Wirhed}},\
  }\href@noop {} {\bibfield  {journal} {\bibinfo  {journal} {Ark. Fysik.}\
  }\textbf {\bibinfo {volume} {\textbf{32}}},\ \bibinfo {pages} {307} (\bibinfo
  {year} {1966})}\BibitemShut {NoStop}%
\bibitem [{\citenamefont {Kopp}\ \emph
  {et~al.}(1966{\natexlab{b}})\citenamefont {Kopp}, \citenamefont
  {Kronekvist},\ and\ \citenamefont {Guntsch}}]{Guntsch1966}%
  \BibitemOpen
  \bibfield  {author} {\bibinfo {author} {\bibfnamefont {I.}~\bibnamefont
  {Kopp}}, \bibinfo {author} {\bibfnamefont {M.}~\bibnamefont {Kronekvist}}, \
  and\ \bibinfo {author} {\bibfnamefont {A.}~\bibnamefont {Guntsch}},\
  }\href@noop {} {\bibfield  {journal} {\bibinfo  {journal} {Ark. Fysik.}\
  }\textbf {\bibinfo {volume} {\textbf{32}}},\ \bibinfo {pages} {371} (\bibinfo
  {year} {1966}{\natexlab{b}})}\BibitemShut {NoStop}%
\bibitem [{\citenamefont {Walker}\ \emph {et~al.}(1993)\citenamefont {Walker},
  \citenamefont {Hedderich},\ and\ \citenamefont {Bernath}}]{Bernath1993}%
  \BibitemOpen
  \bibfield  {author} {\bibinfo {author} {\bibfnamefont {K.~A.}\ \bibnamefont
  {Walker}}, \bibinfo {author} {\bibfnamefont {H.~G.}\ \bibnamefont
  {Hedderich}}, \ and\ \bibinfo {author} {\bibfnamefont {P.~F.}\ \bibnamefont
  {Bernath}},\ }\href@noop {} {\bibfield  {journal} {\bibinfo  {journal} {Mol.
  Phys.}\ }\textbf {\bibinfo {volume} {\textbf{78}}},\ \bibinfo {pages} {577}
  (\bibinfo {year} {1993})}\BibitemShut {NoStop}%
\bibitem [{\citenamefont {Siegbahn}\ \emph {et~al.}(1980)\citenamefont
  {Siegbahn}, \citenamefont {Heiberg}, \citenamefont {Roos},\ and\
  \citenamefont {Levy}}]{Siegbahn1980}%
  \BibitemOpen
  \bibfield  {author} {\bibinfo {author} {\bibfnamefont {P.}~\bibnamefont
  {Siegbahn}}, \bibinfo {author} {\bibfnamefont {A.}~\bibnamefont {Heiberg}},
  \bibinfo {author} {\bibfnamefont {B.}~\bibnamefont {Roos}}, \ and\ \bibinfo
  {author} {\bibfnamefont {B.}~\bibnamefont {Levy}},\ }\href@noop {} {\bibfield
   {journal} {\bibinfo  {journal} {Phys. Scr.}\ }\textbf {\bibinfo {volume}
  {\textbf{21}}},\ \bibinfo {pages} {323} (\bibinfo {year} {1980})}\BibitemShut
  {NoStop}%
\bibitem [{\citenamefont {Werner}\ and\ \citenamefont
  {Knowles}(1985{\natexlab{b}})}]{Werner1985}%
  \BibitemOpen
  \bibfield  {author} {\bibinfo {author} {\bibfnamefont {H.~J.}\ \bibnamefont
  {Werner}}\ and\ \bibinfo {author} {\bibfnamefont {P.~J.}\ \bibnamefont
  {Knowles}},\ }\href@noop {} {\bibfield  {journal} {\bibinfo  {journal} {J.
  Chem. Phys.}\ }\textbf {\bibinfo {volume} {\textbf{82}}},\ \bibinfo {pages}
  {5053} (\bibinfo {year} {1985}{\natexlab{b}})}\BibitemShut {NoStop}%
\bibitem [{\citenamefont {Meissner}(1988)}]{Meissner1988}%
  \BibitemOpen
  \bibfield  {author} {\bibinfo {author} {\bibfnamefont {L.}~\bibnamefont
  {Meissner}},\ }\href@noop {} {\bibfield  {journal} {\bibinfo  {journal}
  {Chem. Phys. Lett.}\ }\textbf {\bibinfo {volume} {\textbf{146}}},\ \bibinfo
  {pages} {204} (\bibinfo {year} {1988})}\BibitemShut {NoStop}%
\bibitem [{\citenamefont {Langhoff}\ and\ \citenamefont
  {Davidison}(1974)}]{Davidson1974}%
  \BibitemOpen
  \bibfield  {author} {\bibinfo {author} {\bibfnamefont {S.}~\bibnamefont
  {Langhoff}}\ and\ \bibinfo {author} {\bibfnamefont {E.~R.}\ \bibnamefont
  {Davidison}},\ }\href@noop {} {\bibfield  {journal} {\bibinfo  {journal}
  {Int. J. Quantum Chemistry}\ }\textbf {\bibinfo {volume} {\textbf{8}}},\
  \bibinfo {pages} {61} (\bibinfo {year} {1974})}\BibitemShut {NoStop}%
\bibitem [{\citenamefont {Peterson}(2015)}]{Peterson2015}%
  \BibitemOpen
  \bibfield  {author} {\bibinfo {author} {\bibfnamefont {K.}~\bibnamefont
  {Peterson}},\ }\href {{http://tyr0.chem.wsu.edu/$\sim$kipeters/basis.html}}
  {\enquote {\bibinfo {title} {{private communication, all the basis sets are
  available from the Peterson group website,
  http://tyr0.chem.wsu.edu/$\sim$kipeters/basis.html save aug-cc-pCVnZ-PP}},}\
  } (\bibinfo {year} {2015})\BibitemShut {NoStop}%
\bibitem [{\citenamefont {Lim}\ \emph {et~al.}(2006)\citenamefont {Lim},
  \citenamefont {Stoll},\ and\ \citenamefont
  {Schwerdtfeger}}]{Schwerdtfeger2006}%
  \BibitemOpen
  \bibfield  {author} {\bibinfo {author} {\bibfnamefont {I.~S.}\ \bibnamefont
  {Lim}}, \bibinfo {author} {\bibfnamefont {H.}~\bibnamefont {Stoll}}, \ and\
  \bibinfo {author} {\bibfnamefont {P.}~\bibnamefont {Schwerdtfeger}},\
  }\href@noop {} {\bibfield  {journal} {\bibinfo  {journal} {J. Chem. Phys.}\
  }\textbf {\bibinfo {volume} {\textbf{124}}},\ \bibinfo {pages} {034107}
  (\bibinfo {year} {2006})}\BibitemShut {NoStop}%
\bibitem [{\citenamefont {Gao}\ and\ \citenamefont {Gao}(2014)}]{Gao2014}%
  \BibitemOpen
  \bibfield  {author} {\bibinfo {author} {\bibfnamefont {Y.}~\bibnamefont
  {Gao}}\ and\ \bibinfo {author} {\bibfnamefont {T.}~\bibnamefont {Gao}},\
  }\href@noop {} {\bibfield  {journal} {\bibinfo  {journal} {Phys. Rev. A}\
  }\textbf {\bibinfo {volume} {\textbf{90}}},\ \bibinfo {pages} {052505}
  (\bibinfo {year} {2014})}\BibitemShut {NoStop}%
\bibitem [{\citenamefont {Jensen}(2005)}]{Jensen2005}%
  \BibitemOpen
  \bibfield  {author} {\bibinfo {author} {\bibfnamefont {F.}~\bibnamefont
  {Jensen}},\ }\href@noop {} {\bibfield  {journal} {\bibinfo  {journal} {Theo.
  Chem. Acc.}\ }\textbf {\bibinfo {volume} {\textbf{113}}},\ \bibinfo {pages}
  {267} (\bibinfo {year} {2005})}\BibitemShut {NoStop}%
\bibitem [{\citenamefont {Karton}\ and\ \citenamefont
  {Martin}(2006)}]{Karton2006}%
  \BibitemOpen
  \bibfield  {author} {\bibinfo {author} {\bibfnamefont {A.}~\bibnamefont
  {Karton}}\ and\ \bibinfo {author} {\bibfnamefont {J.~M.~L.}\ \bibnamefont
  {Martin}},\ }\href@noop {} {\bibfield  {journal} {\bibinfo  {journal} {Theo.
  Chem. Acc.}\ }\textbf {\bibinfo {volume} {\textbf{115}}},\ \bibinfo {pages}
  {330} (\bibinfo {year} {2006})}\BibitemShut {NoStop}%
\bibitem [{\citenamefont {J.\u{C}\'{i}\u{z}ek}(1966)}]{Cizek1966}%
  \BibitemOpen
  \bibfield  {author} {\bibinfo {author} {\bibnamefont {J.\u{C}\'{i}\u{z}ek}},\
  }\href@noop {} {\bibfield  {journal} {\bibinfo  {journal} {J. Chem. Phys.}\
  }\textbf {\bibinfo {volume} {\textbf{45}}},\ \bibinfo {pages} {4256}
  (\bibinfo {year} {1966})}\BibitemShut {NoStop}%
\bibitem [{\citenamefont {{Purvis III}}\ and\ \citenamefont
  {Barlett}(1982)}]{Bartlett1982}%
  \BibitemOpen
  \bibfield  {author} {\bibinfo {author} {\bibfnamefont {G.~D.}\ \bibnamefont
  {{Purvis III}}}\ and\ \bibinfo {author} {\bibfnamefont {R.~J.}\ \bibnamefont
  {Barlett}},\ }\href@noop {} {\bibfield  {journal} {\bibinfo  {journal} {J.
  Chem. Phys.}\ }\textbf {\bibinfo {volume} {\textbf{76}}},\ \bibinfo {pages}
  {1910} (\bibinfo {year} {1982})}\BibitemShut {NoStop}%
\bibitem [{\citenamefont {Watts}\ \emph {et~al.}(1993)\citenamefont {Watts},
  \citenamefont {Gauss},\ and\ \citenamefont {Barlett}}]{Bartlett1993}%
  \BibitemOpen
  \bibfield  {author} {\bibinfo {author} {\bibfnamefont {J.~D.}\ \bibnamefont
  {Watts}}, \bibinfo {author} {\bibfnamefont {J.}~\bibnamefont {Gauss}}, \ and\
  \bibinfo {author} {\bibfnamefont {R.~J.}\ \bibnamefont {Barlett}},\
  }\href@noop {} {\bibfield  {journal} {\bibinfo  {journal} {J. Chem. Phys.}\
  }\textbf {\bibinfo {volume} {\textbf{98}}},\ \bibinfo {pages} {8718}
  (\bibinfo {year} {1993})}\BibitemShut {NoStop}%
\bibitem [{\citenamefont {Wang}\ and\ \citenamefont
  {Andrews}(2004)}]{Andrews2004}%
  \BibitemOpen
  \bibfield  {author} {\bibinfo {author} {\bibfnamefont {X.}~\bibnamefont
  {Wang}}\ and\ \bibinfo {author} {\bibfnamefont {L.}~\bibnamefont {Andrews}},\
  }\href@noop {} {\bibfield  {journal} {\bibinfo  {journal} {J. Phys. Chem. A}\
  }\textbf {\bibinfo {volume} {\textbf{108}}},\ \bibinfo {pages} {11500}
  (\bibinfo {year} {2004})}\BibitemShut {NoStop}%
\bibitem [{\citenamefont {Linstrom}\ and\ \citenamefont
  {Mallard}(2005)}]{NIST2005}%
  \BibitemOpen
  \bibfield  {author} {\bibinfo {author} {\bibfnamefont {P.~J.}\ \bibnamefont
  {Linstrom}}\ and\ \bibinfo {author} {\bibfnamefont {W.~G.}\ \bibnamefont
  {Mallard}},\ }\href {http://webook.nist.gov} {\enquote {\bibinfo {title}
  {{NIST Chemistry WebBook NIST Standard Reference Database Number 69}},}\
  }\bibinfo {howpublished} {{National Institute of Standards and Technology,
  Gaithersburg, MD, USA, http://www.webbook.nist.gov}} (\bibinfo {year}
  {2005})\BibitemShut {NoStop}%
\bibitem [{\citenamefont {Mitroy}\ \emph {et~al.}(2010)\citenamefont {Mitroy},
  \citenamefont {Safronova},\ and\ \citenamefont {Clark}}]{Mitroy2010}%
  \BibitemOpen
  \bibfield  {author} {\bibinfo {author} {\bibfnamefont {J.}~\bibnamefont
  {Mitroy}}, \bibinfo {author} {\bibfnamefont {M.~S.}\ \bibnamefont
  {Safronova}}, \ and\ \bibinfo {author} {\bibfnamefont {C.~W.}\ \bibnamefont
  {Clark}},\ }\href@noop {} {\bibfield  {journal} {\bibinfo  {journal} {J.
  Phys. B: At. Mol. Opt. Phys.}\ }\textbf {\bibinfo {volume} {\textbf{43}}},\
  \bibinfo {pages} {202001} (\bibinfo {year} {2010})}\BibitemShut {NoStop}%
\bibitem [{\citenamefont {C\^ote}\ and\ \citenamefont
  {Dalgarno}(2000)}]{Dalgarno2000}%
  \BibitemOpen
  \bibfield  {author} {\bibinfo {author} {\bibfnamefont {R.}~\bibnamefont
  {C\^ote}}\ and\ \bibinfo {author} {\bibfnamefont {A.}~\bibnamefont
  {Dalgarno}},\ }\href@noop {} {\bibfield  {journal} {\bibinfo  {journal}
  {Phys. Rev}\ }\textbf {\bibinfo {volume} {\textbf{62}}},\ \bibinfo {pages}
  {012709} (\bibinfo {year} {2000})}\BibitemShut {NoStop}%
\bibitem [{\citenamefont {Derevianko}\ \emph {et~al.}(2010)\citenamefont
  {Derevianko}, \citenamefont {Porsev},\ and\ \citenamefont
  {Babb}}]{Derevianko2010}%
  \BibitemOpen
  \bibfield  {author} {\bibinfo {author} {\bibfnamefont {A.}~\bibnamefont
  {Derevianko}}, \bibinfo {author} {\bibfnamefont {S.~G.}\ \bibnamefont
  {Porsev}}, \ and\ \bibinfo {author} {\bibfnamefont {J.~F.}\ \bibnamefont
  {Babb}},\ }\href@noop {} {\bibfield  {journal} {\bibinfo  {journal} {Atomic
  Data and Nuclear Data Tables}\ }\textbf {\bibinfo {volume} {\textbf{96}}},\
  \bibinfo {pages} {323} (\bibinfo {year} {2010})}\BibitemShut {NoStop}%
\bibitem [{\citenamefont {Tang}(1969)}]{Tang1969}%
  \BibitemOpen
  \bibfield  {author} {\bibinfo {author} {\bibfnamefont {K.~T.}\ \bibnamefont
  {Tang}},\ }\href@noop {} {\bibfield  {journal} {\bibinfo  {journal} {Phys.
  Rev.}\ }\textbf {\bibinfo {volume} {\textbf{177}}},\ \bibinfo {pages} {108}
  (\bibinfo {year} {1969})}\BibitemShut {NoStop}%
\bibitem [{\citenamefont {Patil}(2000)}]{Patil2000}%
  \BibitemOpen
  \bibfield  {author} {\bibinfo {author} {\bibfnamefont {S.~H.}\ \bibnamefont
  {Patil}},\ }\href@noop {} {\bibfield  {journal} {\bibinfo  {journal} {Eur.
  Phys. J. D}\ }\textbf {\bibinfo {volume} {\textbf{10}}},\ \bibinfo {pages}
  {341} (\bibinfo {year} {2000})}\BibitemShut {NoStop}%
\bibitem [{\citenamefont {Porsev}\ and\ \citenamefont
  {Derevianko}(2006)}]{Porsev2006}%
  \BibitemOpen
  \bibfield  {author} {\bibinfo {author} {\bibfnamefont {S.~G.}\ \bibnamefont
  {Porsev}}\ and\ \bibinfo {author} {\bibfnamefont {A.}~\bibnamefont
  {Derevianko}},\ }\href@noop {} {\bibfield  {journal} {\bibinfo  {journal}
  {JETP,}\ }\textbf {\bibinfo {volume} {\textbf{102}}},\ \bibinfo {pages} {195}
  (\bibinfo {year} {2006})}\BibitemShut {NoStop}%
\bibitem [{\citenamefont {Patil}\ and\ \citenamefont {Tang}(1997)}]{Patil1997}%
  \BibitemOpen
  \bibfield  {author} {\bibinfo {author} {\bibfnamefont {S.~H.}\ \bibnamefont
  {Patil}}\ and\ \bibinfo {author} {\bibfnamefont {K.~T.}\ \bibnamefont
  {Tang}},\ }\href@noop {} {\bibfield  {journal} {\bibinfo  {journal} {J. Chem.
  Phys.}\ }\textbf {\bibinfo {volume} {\textbf{106}}},\ \bibinfo {pages} {2298}
  (\bibinfo {year} {1997})}\BibitemShut {NoStop}%
\bibitem [{\citenamefont {{Le Roy}}(2014)}]{LeRoy2014}%
  \BibitemOpen
  \bibfield  {author} {\bibinfo {author} {\bibfnamefont {R.~J.}\ \bibnamefont
  {{Le Roy}}},\ }\href@noop {} {\bibfield  {journal} {\bibinfo  {journal}
  {\texttt{LEVEL} 8.2 University of Waterloo Chemical Physics Research Report
  CP-668}\ } (\bibinfo {year} {2014})}\BibitemShut {NoStop}%
\bibitem [{\citenamefont {Ramanaish}\ and\ \citenamefont
  {Lakishman}(1982)}]{Ramanaish1982}%
  \BibitemOpen
  \bibfield  {author} {\bibinfo {author} {\bibfnamefont {M.}~\bibnamefont
  {Ramanaish}}\ and\ \bibinfo {author} {\bibfnamefont {S.}~\bibnamefont
  {Lakishman}},\ }\href@noop {} {\bibfield  {journal} {\bibinfo  {journal}
  {Physica C}\ }\textbf {\bibinfo {volume} {\textbf{113}}},\ \bibinfo {pages}
  {263} (\bibinfo {year} {1982})}\BibitemShut {NoStop}%
\bibitem [{\citenamefont {{Le Roy}}(2013{\natexlab{a}})}]{LeRoy2013}%
  \BibitemOpen
  \bibfield  {author} {\bibinfo {author} {\bibfnamefont {R.~J.}\ \bibnamefont
  {{Le Roy}}},\ }\href@noop {} {\bibfield  {journal} {\bibinfo  {journal}
  {\texttt{betaFIT} University of Waterloo Chemical Physics Research Report
  CP-666}\ } (\bibinfo {year} {2013}{\natexlab{a}})}\BibitemShut {NoStop}%
\bibitem [{\citenamefont {Xie}\ \emph {et~al.}(2014)\citenamefont {Xie},
  \citenamefont {Mishra}, \citenamefont {Kar},\ and\ \citenamefont
  {Xie}}]{Xie2014}%
  \BibitemOpen
  \bibfield  {author} {\bibinfo {author} {\bibfnamefont {J.~C.}\ \bibnamefont
  {Xie}}, \bibinfo {author} {\bibfnamefont {S.~K.}\ \bibnamefont {Mishra}},
  \bibinfo {author} {\bibfnamefont {T.}~\bibnamefont {Kar}}, \ and\ \bibinfo
  {author} {\bibfnamefont {R.-H.}\ \bibnamefont {Xie}},\ }\href@noop {}
  {\bibfield  {journal} {\bibinfo  {journal} {Chem. Phys. Lett.}\ }\textbf
  {\bibinfo {volume} {605}},\ \bibinfo {pages} {137} (\bibinfo {year}
  {2014})}\BibitemShut {NoStop}%
\bibitem [{\citenamefont {Tang}\ and\ \citenamefont
  {Toennies}(1984)}]{Tang1984}%
  \BibitemOpen
  \bibfield  {author} {\bibinfo {author} {\bibfnamefont {K.~T.}\ \bibnamefont
  {Tang}}\ and\ \bibinfo {author} {\bibfnamefont {J.~P.}\ \bibnamefont
  {Toennies}},\ }\href@noop {} {\bibfield  {journal} {\bibinfo  {journal} {J.
  Chem. Phys.}\ }\textbf {\bibinfo {volume} {80}},\ \bibinfo {pages} {3726}
  (\bibinfo {year} {1984})}\BibitemShut {NoStop}%
\bibitem [{\citenamefont {{Le Roy}}\ \emph
  {et~al.}(2006{\natexlab{b}})\citenamefont {{Le Roy}}, \citenamefont {Huang},\
  and\ \citenamefont {Jary}}]{LeRoy2006a}%
  \BibitemOpen
  \bibfield  {author} {\bibinfo {author} {\bibfnamefont {R.~J.}\ \bibnamefont
  {{Le Roy}}}, \bibinfo {author} {\bibfnamefont {Y.~Y.}\ \bibnamefont {Huang}},
  \ and\ \bibinfo {author} {\bibfnamefont {C.}~\bibnamefont {Jary}},\
  }\href@noop {} {\bibfield  {journal} {\bibinfo  {journal} {J. Chem. Phys.}\
  }\textbf {\bibinfo {volume} {125}},\ \bibinfo {pages} {164310} (\bibinfo
  {year} {2006}{\natexlab{b}})}\BibitemShut {NoStop}%
\bibitem [{\citenamefont {Dattani}\ and\ \citenamefont {{Le
  Roy}}(2015)}]{Dattani2015a}%
  \BibitemOpen
  \bibfield  {author} {\bibinfo {author} {\bibfnamefont {N.~S.}\ \bibnamefont
  {Dattani}}\ and\ \bibinfo {author} {\bibfnamefont {R.~J.}\ \bibnamefont {{Le
  Roy}}},\ }\href@noop {} {\bibfield  {journal} {\bibinfo  {journal} {ArXiV}\
  ,\ \bibinfo {pages} {1508.07184}} (\bibinfo {year} {2015})}\BibitemShut
  {NoStop}%
\bibitem [{\citenamefont {Henderson}\ \emph {et~al.}(2013)\citenamefont
  {Henderson}, \citenamefont {Shayesteh}, \citenamefont {Tao}, \citenamefont
  {Haugen}, \citenamefont {Bernath},\ and\ \citenamefont {{Le
  Roy}}}]{Henderson2013}%
  \BibitemOpen
  \bibfield  {author} {\bibinfo {author} {\bibfnamefont {R.}~\bibnamefont
  {Henderson}}, \bibinfo {author} {\bibfnamefont {A.}~\bibnamefont
  {Shayesteh}}, \bibinfo {author} {\bibfnamefont {J.}~\bibnamefont {Tao}},
  \bibinfo {author} {\bibfnamefont {C.}~\bibnamefont {Haugen}}, \bibinfo
  {author} {\bibfnamefont {P.}~\bibnamefont {Bernath}}, \ and\ \bibinfo
  {author} {\bibfnamefont {R.}~\bibnamefont {{Le Roy}}},\ }\href@noop {}
  {\bibfield  {journal} {\bibinfo  {journal} {J. Phys. Chem. A}\ }\textbf
  {\bibinfo {volume} {\textbf{117}}},\ \bibinfo {pages} {133373} (\bibinfo
  {year} {2013})}\BibitemShut {NoStop}%
\bibitem [{\citenamefont {Shayesteh}\ \emph {et~al.}(2007)\citenamefont
  {Shayesteh}, \citenamefont {Henderson}, \citenamefont {{Le Roy}},\ and\
  \citenamefont {Bernath}}]{Shayesteh2007}%
  \BibitemOpen
  \bibfield  {author} {\bibinfo {author} {\bibfnamefont {A.}~\bibnamefont
  {Shayesteh}}, \bibinfo {author} {\bibfnamefont {R.~D.~E.}\ \bibnamefont
  {Henderson}}, \bibinfo {author} {\bibfnamefont {R.~J.}\ \bibnamefont {{Le
  Roy}}}, \ and\ \bibinfo {author} {\bibfnamefont {P.~F.}\ \bibnamefont
  {Bernath}},\ }\href@noop {} {\bibfield  {journal} {\bibinfo  {journal} {J.
  Phys. Chem. A}\ }\textbf {\bibinfo {volume} {\textbf{111}}},\ \bibinfo
  {pages} {12495} (\bibinfo {year} {2007})}\BibitemShut {NoStop}%
\bibitem [{\citenamefont {{Le Roy}}\ \emph {et~al.}(2009)\citenamefont {{Le
  Roy}}, \citenamefont {Dattani}, \citenamefont {Coxon}, \citenamefont {Ross},
  \citenamefont {Crozet},\ and\ \citenamefont {Linton}}]{LeRoy2009}%
  \BibitemOpen
  \bibfield  {author} {\bibinfo {author} {\bibfnamefont {R.~J.}\ \bibnamefont
  {{Le Roy}}}, \bibinfo {author} {\bibfnamefont {N.~S.}\ \bibnamefont
  {Dattani}}, \bibinfo {author} {\bibfnamefont {J.~A.}\ \bibnamefont {Coxon}},
  \bibinfo {author} {\bibfnamefont {A.~J.}\ \bibnamefont {Ross}}, \bibinfo
  {author} {\bibfnamefont {P.}~\bibnamefont {Crozet}}, \ and\ \bibinfo {author}
  {\bibfnamefont {C.}~\bibnamefont {Linton}},\ }\href@noop {} {\bibfield
  {journal} {\bibinfo  {journal} {J. Chem. Phys.}\ }\textbf {\bibinfo {volume}
  {131}},\ \bibinfo {pages} {204309} (\bibinfo {year} {2009})}\BibitemShut
  {NoStop}%
\bibitem [{\citenamefont {{Le Roy}}(2013{\natexlab{b}})}]{LeRoy2013a}%
  \BibitemOpen
  \bibfield  {author} {\bibinfo {author} {\bibfnamefont {R.~J.}\ \bibnamefont
  {{Le Roy}}},\ }\href@noop {} {\bibfield  {journal} {\bibinfo  {journal}
  {\texttt{DPotFit} University of Waterloo Chemical Physics Research Report
  CP-667}\ } (\bibinfo {year} {2013}{\natexlab{b}})}\BibitemShut {NoStop}%
\bibitem [{\citenamefont {Douketis}\ \emph {et~al.}(1982)\citenamefont
  {Douketis}, \citenamefont {Scoles}, \citenamefont {Marchetti}, \citenamefont
  {Zen},\ and\ \citenamefont {Thakkar}}]{Douketis1982}%
  \BibitemOpen
  \bibfield  {author} {\bibinfo {author} {\bibfnamefont {C.}~\bibnamefont
  {Douketis}}, \bibinfo {author} {\bibfnamefont {G.}~\bibnamefont {Scoles}},
  \bibinfo {author} {\bibfnamefont {S.}~\bibnamefont {Marchetti}}, \bibinfo
  {author} {\bibfnamefont {M.}~\bibnamefont {Zen}}, \ and\ \bibinfo {author}
  {\bibfnamefont {A.~J.}\ \bibnamefont {Thakkar}},\ }\href@noop {} {\bibfield
  {journal} {\bibinfo  {journal} {J. Chem. Phys.}\ }\textbf {\bibinfo {volume}
  {76}},\ \bibinfo {pages} {3057} (\bibinfo {year} {1982})}\BibitemShut
  {NoStop}%
\bibitem [{\citenamefont {{Le Roy}}\ and\ \citenamefont
  {Huang}(2002)}]{LeRoy2002}%
  \BibitemOpen
  \bibfield  {author} {\bibinfo {author} {\bibfnamefont {R.~J.}\ \bibnamefont
  {{Le Roy}}}\ and\ \bibinfo {author} {\bibfnamefont {Y.}~\bibnamefont
  {Huang}},\ }\href@noop {} {\bibfield  {journal} {\bibinfo  {journal} {Journal
  of Molecular Structure (Theochem)}\ }\textbf {\bibinfo {volume} {591}},\
  \bibinfo {pages} {175} (\bibinfo {year} {2002})}\BibitemShut {NoStop}%
\bibitem [{\citenamefont {Brown}\ and\ \citenamefont
  {Watson}(1977)}]{Brown1977}%
  \BibitemOpen
  \bibfield  {author} {\bibinfo {author} {\bibfnamefont {J.~M.}\ \bibnamefont
  {Brown}}\ and\ \bibinfo {author} {\bibfnamefont {J.~K.~G.}\ \bibnamefont
  {Watson}},\ }\href@noop {} {\bibfield  {journal} {\bibinfo  {journal} {J.
  Mol. Spectrosc.}\ }\textbf {\bibinfo {volume} {\textbf{65}}},\ \bibinfo
  {pages} {65} (\bibinfo {year} {1977})}\BibitemShut {NoStop}%
\bibitem [{\citenamefont {Semczuk}\ \emph {et~al.}(2013)\citenamefont
  {Semczuk}, \citenamefont {Li}, \citenamefont {Gunton}, \citenamefont {Haw},
  \citenamefont {Dattani}, \citenamefont {Witz}, \citenamefont {Mills},
  \citenamefont {Jones},\ and\ \citenamefont {Madison}}]{Semczuk2013}%
  \BibitemOpen
  \bibfield  {author} {\bibinfo {author} {\bibfnamefont {M.}~\bibnamefont
  {Semczuk}}, \bibinfo {author} {\bibfnamefont {X.}~\bibnamefont {Li}},
  \bibinfo {author} {\bibfnamefont {W.}~\bibnamefont {Gunton}}, \bibinfo
  {author} {\bibfnamefont {M.}~\bibnamefont {Haw}}, \bibinfo {author}
  {\bibfnamefont {N.~S.}\ \bibnamefont {Dattani}}, \bibinfo {author}
  {\bibfnamefont {J.}~\bibnamefont {Witz}}, \bibinfo {author} {\bibfnamefont
  {A.~K.}\ \bibnamefont {Mills}}, \bibinfo {author} {\bibfnamefont {D.~J.}\
  \bibnamefont {Jones}}, \ and\ \bibinfo {author} {\bibfnamefont {K.~W.}\
  \bibnamefont {Madison}},\ }\href@noop {} {\bibfield  {journal} {\bibinfo
  {journal} {Phys. Rev. A}\ ,\ \bibinfo {pages} {052505}} (\bibinfo {year}
  {2013})}\BibitemShut {NoStop}%
\bibitem [{\citenamefont {Bernard}\ \emph {et~al.}(1989)\citenamefont
  {Bernard}, \citenamefont {Effantin}, \citenamefont {d'Incan}, \citenamefont
  {Fabre}, \citenamefont {Stringat},\ and\ \citenamefont
  {Barrow}}]{Bernard1989}%
  \BibitemOpen
  \bibfield  {author} {\bibinfo {author} {\bibfnamefont {A.}~\bibnamefont
  {Bernard}}, \bibinfo {author} {\bibfnamefont {C.}~\bibnamefont {Effantin}},
  \bibinfo {author} {\bibfnamefont {J.}~\bibnamefont {d'Incan}}, \bibinfo
  {author} {\bibfnamefont {G.}~\bibnamefont {Fabre}}, \bibinfo {author}
  {\bibfnamefont {R.}~\bibnamefont {Stringat}}, \ and\ \bibinfo {author}
  {\bibfnamefont {R.~F.}\ \bibnamefont {Barrow}},\ }\href@noop {} {\bibfield
  {journal} {\bibinfo  {journal} {Mol. Phys.}\ }\textbf {\bibinfo {volume}
  {\textbf{67}}},\ \bibinfo {pages} {1} (\bibinfo {year} {1989})}\BibitemShut
  {NoStop}%
\end{thebibliography}%

\end{document}